\documentclass[aps,prb,reprint,superscriptaddress]{revtex4-1}
\usepackage{amssymb,amsmath,mathptmx}
\usepackage{graphicx}
\usepackage{dcolumn}
\usepackage{bm}
\usepackage{hyperref}
\usepackage{xcolor}
\usepackage[utf8x]{inputenc}
\usepackage{array}
\newcolumntype{C}{>{\centering\arraybackslash}p{1em}}
\newcommand{\kbf}{\mathbf{k}}

\newcommand{\be}{\begin{equation}}
\newcommand{\ee}{\end{equation}}
\newcommand{\bea}{\begin{eqnarray}}
\newcommand{\eea}{\end{eqnarray}}
\begin{document}


\title{Inversion in a four-terminal superconducting device on the
  quartet line:\\ II. Quantum dot and Floquet theory}

\author{R\'egis M\'elin}

\affiliation{Univ. Grenoble-Alpes, CNRS, Grenoble INP\thanks{Institute
    of Engineering Univ. Grenoble Alpes}, Institut NEEL, 38000
  Grenoble, France}

\author{Beno\^{\i}t Dou\c{c}ot}

\affiliation{Laboratoire de Physique Théorique et Hautes Energies,
  Sorbonne Universit\'e and CNRS UMR 7589, 4 place Jussieu, 75252
  Paris Cedex 05, France}

\begin{abstract}
  In this paper, we consider a quantum dot connected to four
  superconducting terminals biased at opposite voltages on the quartet
  line. The grounded superconductor contains a loop threaded by the
  magnetic flux $\Phi$. We provide Keldysh microscopic calculations
  and physical pictures for the voltage-$V$ dependence of the quartet
  current. Superconductivity is expected to be stronger at
  $\Phi/\Phi_0=0$ than at $\Phi/\Phi_0=1/2$. However, inversion
  $I_{q,c}(V,0)<I_{q,c}(V,1/2)$ is obtained in the critical current
  $I_{q,c}(V,\Phi/\Phi_0)$ on the quartet line in the voltage-$V$
  ranges which match avoided crossings in the Floquet spectrum at
  $(V,\Phi/\Phi_0=0)$ but not at $(V,1/2)$. A reduction in $I_{q,c}$
  appears in the vicinity of those avoided crossings, where
  Landau-Zener tunneling produces dynamical quantum mechanical
  superpositions of the Andreev bound states.  In addition, $\pi$-$0$
  and $0$-$\pi$ cross-overs emerge in the current-phase relations as
  $V$ is further increased. The voltage-induced $\pi$-shift is
  interpreted as originating from the nonequilibrium Floquet
  populations produced by voltage biasing. The numerical calculations
  reveal that the inversion is robust against strong Landau-Zener
  tunneling and many levels in the quantum dot. Our theory provides a
  simple ``Floquet level and population'' mechanism for inversion
  tuned by the bias voltage~$V$, which paves the way towards more
  realistic models for the recent Harvard group experiment where the
  inversion is observed.
\end{abstract}

\maketitle

\section{Introduction}

Quantum optics and cold atom experiments revealed entanglement among
two \cite{EPR,Bell,Aspect}, three \cite{GHZ-theorie,GHZ-3photons} or
four \cite{Wineland} particles. The progress in nanofabrication
technology made it possible to consider solid-state analogues since
the early 2000s. {However, twenty years} after the first theoretical
and experimental efforts (see for instance
Refs.~\onlinecite{theory-CPBS1,theory-CPBS2,theory-CPBS3,theory-CPBS4,theory-CPBS4-bis,theory-CPBS4-ter,theory-CPBS5,theory-CPBS6,theory-CPBS7,theory-CPBS8,theory-CPBS9,theory-CPBS10,theory-CPBS11}
for the theory, and
Refs.~\onlinecite{exp-CPBS1,exp-CPBS2,exp-CPBS3,exp-CPBS4,exp-CPBS5,exp-CPBS6,exp-CPBS7,exp-CPBS8}
for the experiments), no proof of entanglement between pairs of
electrons has been reported so far in solid-state superconducting
nanoscale electronic {devices.  Instead, solid-state experiments}
\cite{exp-CPBS1,exp-CPBS2,exp-CPBS3,exp-CPBS4,exp-CPBS5,exp-CPBS6,exp-CPBS7,exp-CPBS8}
provided evidence for correlations among pairs of electrons in
three-terminal ferromagnet-superconductor-ferromagnet ($F_aSF_b$) or
normal metal-superconductor-normal metal ($N_aSN_b$) devices. For
instance, measurements of the nonlocal conductance ${\cal
  G}_{a,b}=\partial I_a/\partial V_b$ demonstrated
\cite{exp-CPBS1,exp-CPBS2,exp-CPBS3,exp-CPBS4,exp-CPBS5,exp-CPBS6,exp-CPBS7,exp-CPBS8}
how the current $I_a$ through lead $F_a$ or $N_a$ depends on the
voltage $V_b$ on lead $F_b$ or $N_b${, the superconductor $S$ being
  grounded}. In addition, the zero-frequency positive current-current
cross-correlations $S_{a,b}$ in three-terminal $N_aSN_b$ beam
splitters demonstrated \cite{exp-CPBS7,exp-CPBS8} the theoretically
predicted
\cite{theory-noise1,theory-noise2,theory-noise3,theory-noise4,theory-noise5,theory-noise6,theory-noise7,theory-noise8,theory-noise9,theory-noise10,theory-noise11,theory-noise12}
quantum fluctuations of the current operators $\hat{I}_a$ and
$\hat{I}_b$.

{The nonstandard quantum mechanical exchange of ``the quartets''
  \cite{theorie-quartets1,theorie-quartets1bis} is operational in
  $(S_a,S_b,S_c)$ three-terminal Josephson junctions, which realize
  all-superconducting analogues of the above mentioned $N_aSN_b$ and
  $F_aSF_b$ three-terminal Cooper pair beam splitters. These quartets
  involve transient correlations among four fermions: they take two
  ``incoming'' pairs from $S_a$ and $S_b$ biased at $\pm V$, and
  transmit the ``outgoing'' ones into the grounded $S_c$ after
  exchanging partners. As shown in
  Refs.~\onlinecite{theorie-quartets1,theorie-quartets1bis}, energy
  conservation implies that the quartets can be revealed as
  dc-Josephson anomaly on the so-called ``quartet line'' $V_a+V_b=0$
  in the $(V_a,V_b)$ voltage plane, with $V_c=0$ for the grounded
  $S_c$. Further developments including Floquet theory and zero- and
  finite-frequency noise calculations are provided in
  Refs.~\onlinecite{theorie-quartets2,theorie-quartets3,theorie-quartets5,theorie-quartets6,theorie-quartets7,theorie-quartets8}.}

{Experimental evidence for the quartet Josephson
  anomaly was published by two groups:}

{(i) The Grenoble group \cite{Lefloch} reported the quartet anomaly in
  three-terminal Aluminum/Copper Josephson junctions \cite{Lefloch},
  where the experimental data for elements of the dc-nonlocal
  resistance matrix are color-plotted in the $(V_a,V_b)$ voltage
  plane.}

{(ii) The Weizmann Institute group \cite{Heiblum} confirmed the
  Josephson-like quartet anomaly with three-terminal Josephson
  junctions connecting a semiconducting nanowire. In addition,
  Ref.~\onlinecite{Heiblum} presents measurements of the
  current-current cross-correlations, interpreted as the quantum
  fluctuations in the quartet current originating from Landau-Zener
  tunneling between the branches of Andreev bound states (ABS). The
  dynamics of the phases is set by the Josephson relation
  $\varphi_a(t)=2eVt/\hbar+\varphi_a$,
  $\varphi_b(t)=2eVt/\hbar+\varphi_b$ and $\varphi_c(t)=\varphi_c$ for
  $(S_a,S_b,S_c)$ biased at $(V,-V,0)$ respectively. Regarding the
  quartets in three-terminal Josephson junctions, the predicted
  \cite{theorie-quartets5} and the measured \cite{Heiblum} positive
  cross-correlations $S_{a,b}>0$ turn out to be in a qualitative
  agreement with each other. The cross-correlations $S_{a,b}>0$ are
  indeed expected to be generically positive, as for any splitting
  process such as Cooper pair splitting
  \cite{theory-noise8,theory-noise9,theory-noise11}. }

\begin{figure}[htb]
  \includegraphics[width=.8\columnwidth]{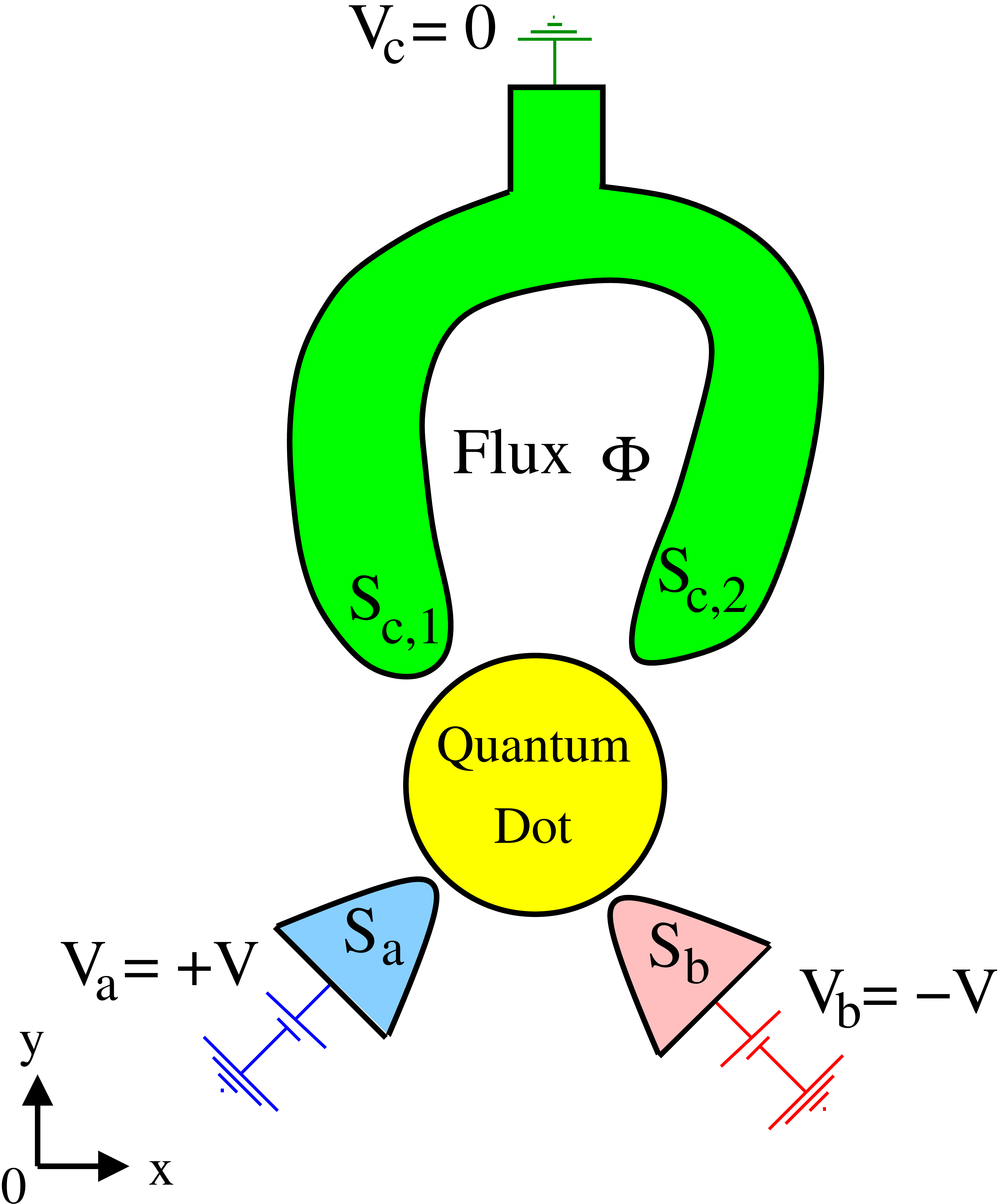}
  \caption{{\it The considered four-terminal device:} Four
    superconducting contacts $S_a$, $S_b$, $S_{c,1}$ and $S_{c,2}$ are
    connected to a quantum dot. The leads $S_a$ and $S_b$ are biased
    at $\pm V$, and $S_{c,1}$, $S_{c,2}$ belong to the same grounded
    terminal $S_c$ to which is connected a loop pierced by the flux
    $\Phi$. The quantum dot has a single level at zero energy, except
    in section~\ref{sec:num2} dealing with a multilevel quantum
    dot.\label{fig:device}}
\end{figure}

{A third experiment realized recently in the Harvard group
  \cite{Harvard-group-experiment} deals with a four-terminal
  $(S_a,S_b,S_{c,1},S_{c,2})$ Josephson junction containing a loop
  pierced by the flux $\Phi$ and biased at
  $(V_a,V_b,V_{c,1},V_{c,2})=(V,-V,0,0)$. Namely, the grounded loop is
  terminated by the contact points $S_{c,1}$ and $S_{c,2}$ and the
  superconducting leads $S_a$ and $S_b$ which do not contain loops are
  biased at $V_{a,b}=\pm V$ respectively, see figure~\ref{fig:device}.
  The recent Harvard group experiment \cite{Harvard-group-experiment}
  features the additional control parameter of the reduced flux
  $\Phi/\Phi_0$, which was not there in the previous Grenoble
  \cite{Lefloch} and Weizmann Institute \cite{Heiblum} group
  experiments.}

{The Harvard group \cite{Harvard-group-experiment}
  reports dc-Josephson anomaly along the ``quartet line'' $V_a+V_b=0$
  (with $V_c=0$ for the grounded $S_c$), which confirms the preceding
  Grenoble \cite{Lefloch} and Weizmann Institute \cite{Heiblum} group
  experiments. In addition, the Harvard group data show that the
  quartet critical current $I_{q,c}(V,\Phi/\Phi_0)$ nontrivially
  depends on both values of the voltage $V$ and the reduced flux
  $\Phi/\Phi_0$, {\it i.e.} inversion $I_{q,c}(V,1/2)>I_{q,c}(V,0)$ is
  observed in a given voltage window, even if, at first glance,
  superconductivity should be stronger at $\Phi/\Phi_0=0$ than at
  $\Phi/\Phi_0=1/2$. The Harvard group experiment
  \cite{Harvard-group-experiment} challenges the theory of the
  quartets
  \cite{theorie-quartets1,theorie-quartets1bis}
  with respect to mechanisms for the inversion between
  $\Phi/\Phi_0=0$ and $\Phi/\Phi_0=1/2$.}

{It was shown in the preceding  \cite{paperI} paper~I
  that inversion in $I_{q,c}(\Phi/\Phi_0)$ between $\Phi/\Phi_0=0$ and
  $\Phi/\Phi_0=1/2$, {\it i.e.} $I_{q,c}(0)<I_{q,c}(1/2)$, can result
  from interference between the three-terminal quartets and the
  four-terminal split quartets if a 2D metal connects the four
  superconductors. Namely, perturbation theory in the tunnel
  amplitudes combined to the $V=0^+$ adiabatic limit yield
  $\pi$-shifted three-terminal and $0$-shifted four-terminal quartets,
  which automatically implies ``inversion between $\Phi/\Phi_0=0$ and
  $\Phi/\Phi_0=1/2$''.}

{Major difference appears between the preceding paper~I and the
  following paper~II: 2D metal is connected to four superconducting
  lead in paper~I\cite{paperI} whereas the following paper~II
  considers 0D quantum dot which is is not directly realized in the
  Harvard group \cite{Harvard-group-experiment}, given the large
  dimension of the graphene sheet in this experiment
  \cite{Harvard-group-experiment}. Nevertheless, simple models are
  often useful and the following paper~II provides useful theoretical
  input on how inversion can result from changing the bias
  voltage~$V$. Paper~III will discuss whether the physical picture of
  the following paper~II can extrapolate to the 2D metal of paper~I
  \cite{paperI}.}

The paper is organized as the
following. {Section~\ref{sec:summary-of-the-paper} presents a summary
  of the main results of the paper.}  The model and the Hamiltonians
are provided in section~\ref{sec:model-and-Hamiltonian}. The rate of
Landau-Zener tunneling is evaluated in section~\ref{sec:R}, in
connection with the Keldysh numerical calculations of
section~\ref{sec:num1}.  Section~\ref{sec:num2} presents robustness of
the inversion against changing the coupling parameters for a single
level quantum dot, and against multichannel effects. Concluding
remarks are presented in section~\ref{sec:conclusions}.

\section{Summary of the main results}
\label{sec:summary-of-the-paper}

This section presents a connection to the known physics of Multiple
Andreev Reflections (MAR) (see section~\ref{sec:MAR}) and a summary of
the main results of this paper~II (see
section~\ref{sec:mechanism-inversion}).

\subsection{Connection with Multiple Andreev Reflections
  (MAR)}
\label{sec:MAR}

{Dissipationless dc-Josephson current \cite{Josephson} carried by the
  ABS \cite{Andreev} flows across a two-terminal weak link
  \cite{Likharev} connecting the superconductors $S_1$ and $S_2$ in
  the presence of phase biasing at $\varphi_{2T} = \varphi_2 -
  \varphi_1\ne 0$ and vanishingly small voltage drop $V_{2T} = V_2 -
  V_1 = 0$. The Josephson effect has applications to superconducting
  quantum interference devices used {\it e.g.} for quantum information
  processing \cite{Kouznetsov,Clarke1,Clarke2,Devoret}. A number of
  experiments provided direct evidence for the ABS, see for instance
  Refs.~\onlinecite{Janvier,Pillet,Dirks,Bretheau1,Bretheau2}.}

{Biasing a superconducting weak link at voltage
  $V_{2T}=V_2-V_1\ne 0$ produces dc-current of MAR at subgap voltage
  $eV_{2T} < 2 \Delta$.}  {Break-junction experiments
  \cite{Scheer} observed the predicted \cite{Averin,Cuevas}
  dc-current-voltage characteristics of the MAR. In addition,
  excellent agreement was obtained between the voltage dependence of
  the zero-frequency quantum noise \cite{Cron} and the calculated Fano
  factor \cite{Cuevas3}.}

{Regarding the MAR, the following situations turn out
  to be drastically different:}

{(i) First, the superconducting weak link bridging
  $S_1$ and $S_2$ is described by a single hopping amplitude in
  Refs.~\onlinecite{Averin,Cuevas}.}

{(ii) Second, a quantum dot with single level at zero energy is
  considered in the following paper.}

{Concerning the above item~(i), the equilibrium ABS
  plotted as a function of the phase difference $\Delta \varphi=
  \varphi_2-\varphi_1$ necessarily touch the continua at the energies
  $\pm \Delta$ if $\Delta \varphi = 0$. At finite bias voltage $V$,
  the phase difference $\varphi_{2T}(t)=2eVt/\hbar+\varphi_{2T}(0)$
  is linear in time, and $\Delta \varphi(t)=2\pi n$ is realized
  periodically, with $n$ an integer. This produces strong coupling of
  the ABS to the quasiparticle continua, resulting in the smooth
  energy-dependence of the spectral currents reported in
  Ref.~\onlinecite{Cuevas}.}

{Now, if a quantum dot connects two superconductors $S_1$ and $S_2$
  according to the above item~(ii), then the ABS do not touch the
  superconducting gap edge singularities at any value of
  $\varphi_{2T}$. Instead, at zero phase difference, the ABS have
  typical energy set by the normal-state line-width broadening
  $\Gamma$. In the following calculations, the values of the $\Gamma$s
  are taken as being smaller than the superconducting gap $\Delta$,
  thus the ABS touch $\pm \Delta$ neither at $\varphi_{2T}=0$ nor at
  arbitrary $\varphi_{2T}$.}

{Considering now biasing at finite voltage~$V$ for the quantum dot in
  the above item~(ii), the energy gap between the maximal ABS energy
  and the gap edge singularity at $\Delta$ implies protection with
  respect to relaxation due to direct coupling to the continua. Then,
  the spectral currents feature a sequence of narrow resonances within
  the energy window of the gap, see the forthcoming
  figure~\ref{fig:spectral_currents}~a2-d2. The energy/frequency
  dependence of the spectral current on
  figure~\ref{fig:spectral_currents}~a2-d2 for the quantum dot in the
  above item~(ii) differs drastically from the smooth variations of
  the spectral current relevant to the item~(i), see
  Ref.~\onlinecite{Cuevas}.}  {Given these observations, the quantum
  dot connecting two superconductors according to the above item~(ii)
  can legitimately be considered as being relevant to ``Floquet
  theory'', and the terms ``Floquet levels'' and ``Floquet
  populations'' can be used.}

In a three-terminal device, the so-called quartets refer to the
microscopic quantum process of two Cooper pairs from $S_a$ and $S_b$
biased at $\pm V$, which exchange partners and transmit the outgoing
pairs into the grounded $S_c$. {The terms ``quartet phase'', ``quartet
  line'' and ``quartet critical currents'' are used beyond
  perturbation theory in the tunneling amplitudes as a convenient
  wording. }

{Finally, we note that our model is strictly speaking 0D, {\it i.e.}
  the quantum dot consists of a single tight-binding site. But this
  ``0D quantum dot'' holds more generally for experimental devices
  fabricated with ``quasi-0D'' quantum dots having energy level
  spacing $\delta_{dot}\gg \Delta$ which is much larger than the
  superconducting gap $\Delta$, but $\delta_{dot}\ll W$ is small
  compared to the band-width $W$. Said differently, our calculations
  capture ``quasi-0D'' quantum dots with linear dimension which is
  large compared to the Fermi wave-length but small compared to the
  BCS coherence length.}

\begin{figure}[htb]
  \includegraphics[width=.7\columnwidth]{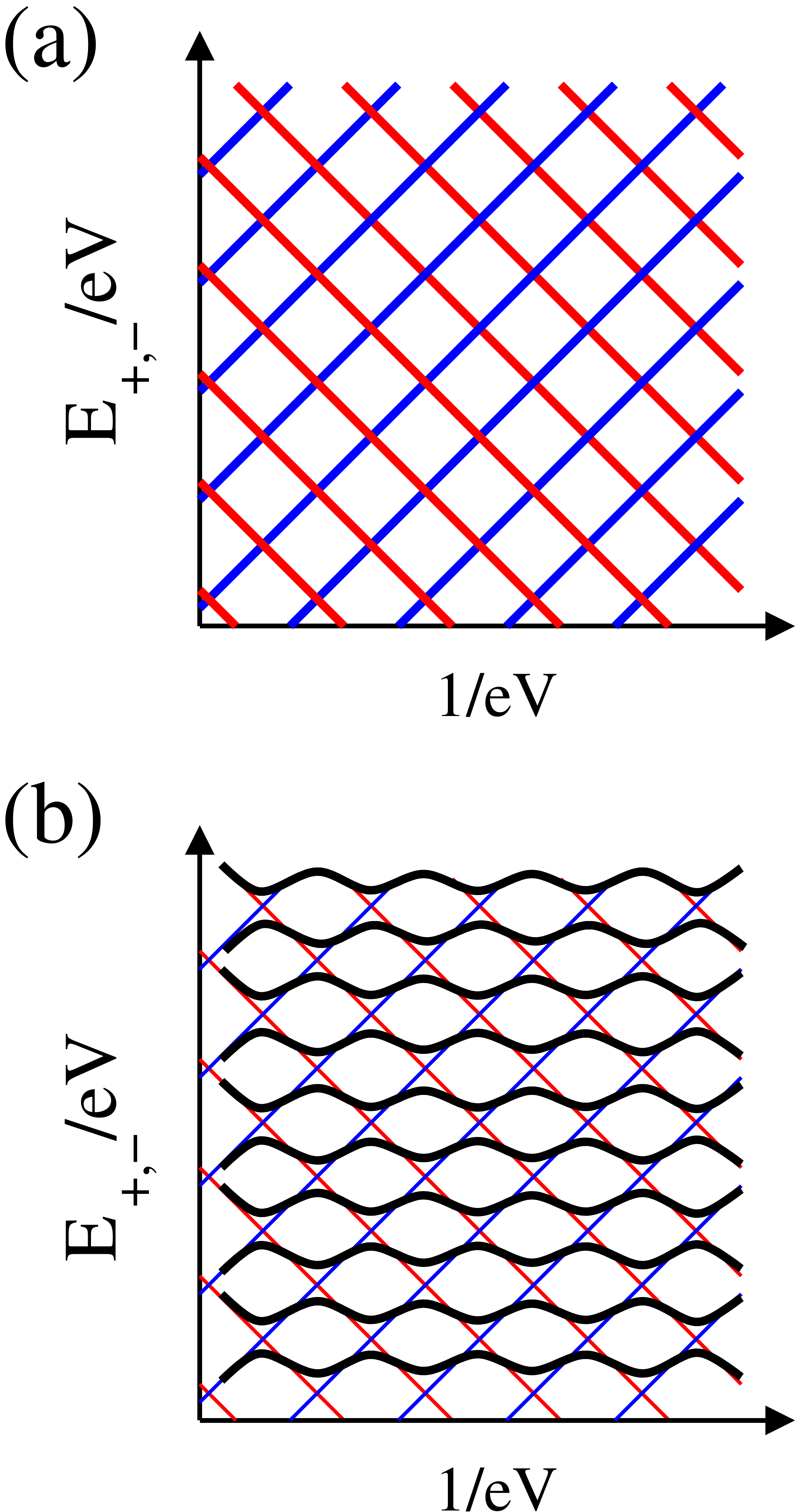}
  \caption{{{\it Schematics of the Floquet spectra:} Panel a shows the
      Floquet spectrum in the classical limit, using the
      $(1/eV,E_{+,-}/eV)$ scaling plots, see
      Eqs.~(\ref{eq:E-plus-2})-(\ref{eq:E-minus-2}).  The
      $1/eV$-dependence of $E_{+,-}/eV$ are shown on panel a by the
      blue upwards and red downwards solid lines respectively.  Panel
      b shows the Floquet spectra in the presence of weak Landau-Zener
      tunneling, which produces avoided crossings. The Floquet spectra
      are shown by the black lines on panel b.}
    \label{fig:supplementary-1}}
\end{figure}

\begin{figure*}[htb]
  \includegraphics[width=.7\textwidth]{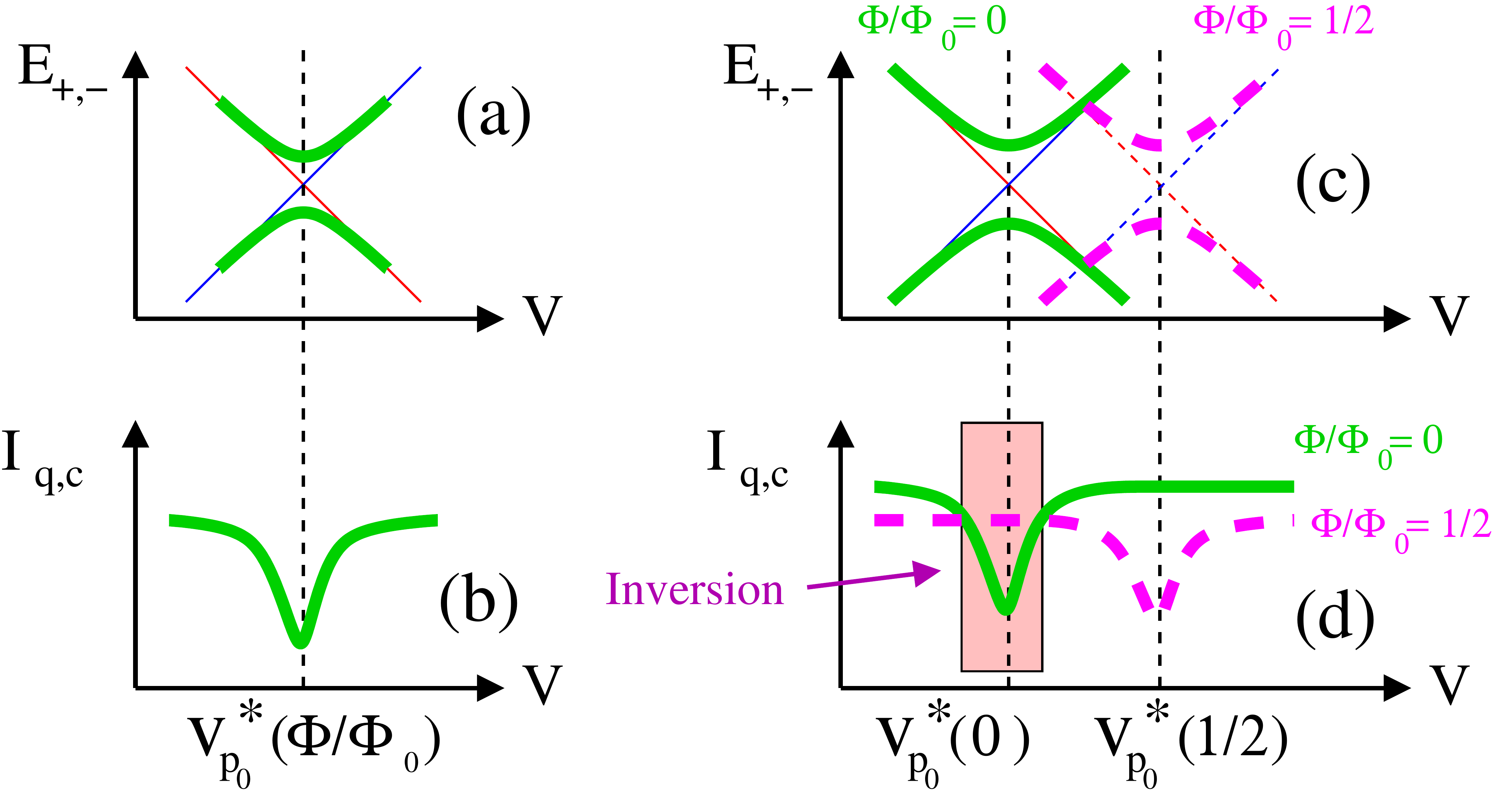}
  \caption{{{\it The mechanism leading to the
        inversion:} At fixed reduced flux $\Phi/\Phi_0$, the avoided
      crossing between two Floquet levels plotted as a function of the
      voltage $V$ on $x$-axis (panel a) is accompanied by a ``dip'' in
      the quartet critical current $I_{q,c}$ (panel b). Panels c and d
      compare the Floquet spectra to $I_{q,c}$ between $\Phi/\Phi_0=0$
      and $\Phi/\Phi_0=1/2$. Panels c and d show schematically how
      ``inversion'' can be produced in $I_{q,c}(V,\Phi/\Phi_0)$, {\it
        i.e.}  $I_{q,c}(V,0)<I_{q,c}(V,1/2)$ in the voltage window
      shown on the figure.}
    \label{fig:supplementary-2}}
\end{figure*}

\subsection{Summary of the main results}
\label{sec:mechanism-inversion}

{Now, we summarize the main results of this paper~II, starting in
  section~\ref{sec:mechanism-weak} with the simple limits of weak
  Landau-Zener tunneling and quantum dot with a single level at zero
  energy. Section~\ref{sec:mechanism-generalization} introduces our
  numerical results for strong Landau-Zener tunneling and multilevel
  quantum dots, {\it i.e.} beyond the single level 0D quantum dot in
  the limit of weak Landau-Zener tunneling.}

\subsubsection{A simple mechanism for the inversion
    at weak Landau-Zener tunneling}

\label{sec:mechanism-weak}

{We start with discussing weak Landau-Zener tunneling for a
  single-level quantum dot having a level at zero
  energy. Specifically, we introduce a connection between two sides of
  the problem:}

{(a) Inversion in $I_{q,c}(V,\Phi/\Phi_0)$ between
    $\Phi/\Phi_0=0$ and $\Phi/\Phi_0=1/2$, {\it i.e.}
    $I_{q,c}(V,0)<I_{q,c}(V,1/2)$.}

{(b) The presence/absence of avoided crossings
  in the Floquet spectra at $\Phi/\Phi_0=0$ and $\Phi/\Phi_0=1/2$
  respectively.}

{Generally speaking, in absence of bias voltage, any equilibrium
  quantum mechanical Hamiltonian can be decomposed into independent
  blocks once the symmetries have been taken into account. Within each
  block, the energy levels plotted as a function of parameters show
  avoided crossings and repulsion.}  {We note that avoided crossings
  in Floquet spectra appeared previously in the literature, see for
  instance Refs.~\onlinecite{Bentosela,Arevalo,Eckardt,Hone}.}

{The dc-Josephson effect is classical in the $V=0$ or $V=0^+$
  equilibrium or adiabatic limits. The classical approximation to the
  finite-$V$ Floquet spectrum is the following \cite{theorie-quartets7}:
\begin{eqnarray}
  \label{eq:E-plus}
  E_{+,p} &=& \langle E_{ABS} \rangle_k + 2 p eV\\
  E_{-,q} &=& - \langle E_{ABS} \rangle_k + 2 q eV
  \label{eq:E-minus}
  .
\end{eqnarray}
where $p$ and $q$ are two integers, and the average $\langle E_{ABS}
\rangle_k$ of the (positive) ABS energy $E_{ABS}$ is taken over the
fast phase variable parameterized by the variable $k$:
\begin{eqnarray}
  \label{eq:k1}
  \varphi_a(k)&=&\varphi_a+k\\
  \varphi_b(k)&=&\varphi_b-k\\
\varphi_{c,1}(k)&=&\varphi_{c,1}\\
\varphi_{c,2}(k)&=&\varphi_{c,2}
\label{eq:k4}
,
\end{eqnarray}
where $(\varphi_a,\varphi_b,\varphi_{c,1},\varphi_{c,2})$ are the
phases of $(S_a,S_b,S_{c,1},S_{c,2})$ respectively. The variable $k$
in Eqs.~(\ref{eq:k1})-(\ref{eq:k4}) stands for
\begin{equation}
  \label{eq:k-def}
  k=\frac{2eVt}{\hbar}
  .
\end{equation}
Eqs.~(\ref{eq:E-plus})-(\ref{eq:E-minus}) are demonstrated from
Bohr-Sommerfeld quantization in Ref.~\onlinecite{theorie-quartets7}.
They receive the simple interpretation that, classically, the Floquet
spectra correspond to adding or subtracting multiples of the voltage
energy $2 eV$ to the adiabatic-limit ABS energies $\langle E_{ABS}
\rangle_k$, where $\pm 2eV$ is the energy for transferring a Cooper
pair between the grounded $S_{c,1}$, $S_{c,2}$ and $S_{a,b}$ biased at
$\pm V$.}

{The classical approximation to the Floquet
  spectrum given by Eqs.~(\ref{eq:E-plus})-(\ref{eq:E-minus}) yields
\begin{eqnarray}
  \label{eq:E-plus-2}
  \frac{E_{+,p}}{eV} &=& \frac{\langle E_{ABS} \rangle_k}{eV} + 2 p\\
  \frac{E_{-,q}}{eV} &=& - \frac{\langle E_{ABS} \rangle_k}{eV} + 2 q
  \label{eq:E-minus-2}
  .
\end{eqnarray}
Figure~\ref{fig:supplementary-1}a shows schematically $E_{+,p}/eV$ and
$E_{-,q}/eV$ as a function of the inverse voltage $1/eV$, according to
Eqs.~(\ref{eq:E-plus-2})-(\ref{eq:E-minus-2}). This
$(1/eV,E_{\pm,p}/eV)$ scaling \cite{Bentosela} yields regular pattern
of the Floquet levels.}

{Eqs.~(\ref{eq:E-plus})-(\ref{eq:E-minus}) and
  Eqs.~(\ref{eq:E-plus-2})-(\ref{eq:E-minus-2}) imply the sequence of
  voltages $\{V_{cross,n}\}$ of the nonavoided crossings on
  figure~\ref{fig:supplementary-1}a. The values of $\{V_{cross,n}\}$
  are such that $e V_{cross,n} = \langle E_{ABS} \rangle_k /n$, with
  $n = q - p$ corresponding to $E_{+,p}(V_{cross,n}) =
  E_{-,q}(V_{cross,n})$, see figure~\ref{fig:supplementary-1}a. }

{ Landau-Zener tunneling between the ABS introduces
  quantum mechanical effects, as if the Planck constant was
  proportional to the bias voltage. This results in opening
  gaps in the Floquet spectra, see the schematic
  figure~\ref{fig:supplementary-1}b where the scaling variables given
  by Eqs.~(\ref{eq:E-plus-2})-(\ref{eq:E-minus-2}) are used.}

At this point, we further comment on the connection to MAR in
two-terminal Josephson junctions. Refs. \onlinecite{Averin,Cuevas}
demonstrate that the adiabatic limit is solely realized at very low
voltage~$V$ if a hopping amplitude connects two superconductors. But
if the weak link consists of a quantum dot, then adiabaticity is
obtained in windows of the bias voltage~$V$ which is ``in between''
consecutive avoided crossings in the Floquet spectrum plotted as a
function of~$V$. This implies adiabaticity at much higher voltage if a
quantum dot is used instead of the hopping hopping amplitude relevant
to break junction experiments.

{Considering now the wave-function, the Floquet-Bogoliubov-de Gennes
  wave-function is a quantum superposition between the negative and
  the positive-energy ABS if the voltage~$V$ and the reduced flux
  $\Phi/\Phi_0$ are tuned at avoided crossings in the Floquet
  spectrum. The two ABS carry opposite currents at equilibrium and
  thus, ``quantum superposition between the ABS'' reduces the quartet
  current.}

{Thus, weak Landau-Zener tunneling implies the items~(a) and~(b) at
  the beginning of this subsection~\ref{sec:mechanism-inversion}, {\it
    i.e.} avoided crossings in the Floquet spectrum are accompanied by
  dips in the quartet critical current, see
  figure~\ref{fig:supplementary-2}~a-b.} We note that the Floquet
spectra appearing in figures~\ref{fig:low_voltage_window1}~a and c are
shown schematically in a restricted energy interval on the $y$-axis,
in comparison with the forthcoming
figure~\ref{fig:low_voltage_window1} showing a larger energy interval
for the reduced Floquet energies $E_n/eV$.

{Going one step further, we argue now that inversion can be produced
  between $\Phi/\Phi_0=0$ and $\Phi/\Phi_0=1/2$ in the quartet
  critical current $I_{q,c}(V,\Phi/\Phi_0)$, {\it i.e.}
  $I_{q,c}(V,0)<I_{q,c}(V,1/2)$.}
  
{Namely, we envision that plotting the Floquet spectra as a function
  of the voltage $V$ produces the sequence $\{V^*_p(\Phi/\Phi_0)\}$ of
  the $V$-values at the avoided crossings, see
  figure~\ref{fig:supplementary-2}c. ``Avoided crossings in the
  Floquet spectrum at $\Phi/\Phi_0=0$'' for $V\simeq V^*_{p_0}(0)$ are
  in general not accompanied by ``avoided crossing at
  $\Phi/\Phi_0=1/2$'' at the same $V\simeq V^*_{p_0}(0)$, see
  figure~\ref{fig:supplementary-2}c. Then, the quartet current can be
  significantly reduced at $\Phi/\Phi_0=0$ but not at
  $\Phi/\Phi_0=1/2$, see figures~\ref{fig:supplementary-2}~c-d. This
  shows that ``hybridization between the ABS'' can produce inversion
  in $I_{q,c}(V,\Phi/\Phi_0)$ between $\Phi/\Phi_0=0$ and
  $\Phi/\Phi_0=1/2$ in the simple limit of single-level quantum dot
  with weak Landau-Zener tunneling. This ``scenario'' is put to the
  test of numerical calculations in the forthcoming
  section~\ref{sec:num1}.}

\subsubsection{Beyond weak Landau-Zener tunneling and single-level
  quantum dot}
\label{sec:mechanism-generalization}
{The paper presents in section~\ref{sec:num2} the following additional
  results:}

{(i) The connection between the extrema in the Floquet
  spectrum and the minima in the quartet critical current holds more
  generally for strong Landau-Zener tunneling, see figures~3 and~4 in
  the Supplemental Material\cite{supplemental-material}.}

{(ii) The inversion appears generically for a multilevel quantum dot,
  see the forthcoming
  figure~\ref{fig:inversion-multilevel-quantum-dot-1} in the paper.}

{(iii) We provide evidence for $\pi$-shifted quartet current-quartet
  phase relations in narrow voltage windows, which are interpreted in
  terms of the nontrivial Floquet populations produced at moderately
  large bias voltage, see section~\ref{sec:pi-shift}.}


\section{Model and Hamiltonians}
\label{sec:model-and-Hamiltonian}

In this section, we present the model and the
Hamiltonians. Specifically, the single level quantum dot device
Hamiltonian is presented in section~\ref{sec:single-dot}. The infinite
gap limit and the gauge-invariant quartet phase variable are presented
in section~\ref{sec:infinite-gap}. The expression of the quartet
current is provided in section~\ref{sec:quartet-current}. The
parameters used in the numerical calculation are given in
section~\ref{sec:parameters}. The multilevel quantum dot is presented
in subsection~\ref{sec:multilevel-quantum-dot-model} and inversion in
the $V=0^+$ adiabatic limit is discussed in
subsection~\ref{sec:adiabatic}.

\subsection{Single-level quantum dot}
\label{sec:single-dot}
In this subsection, we provide the Hamiltonian of the four-terminal
device in figure~\ref{fig:device}, in the limit where the quantum dot
supports a single level at zero energy.

The Hamiltonian is the sum of the BCS Hamiltonian of the
superconducting leads and the tunneling term between the dot and the
leads. In absence of voltage biasing, the Hamiltonian of each
superconducting lead takes the form
\begin{eqnarray}
  \label{eq:H-BCS1}
        {\cal H}_{BCS}&=&-W \sum_{\langle i,j \rangle}
        \sum_{\sigma=\uparrow,\downarrow} \left(c_{i,\sigma}^+
        c_{j,\sigma}+ c_{j,\sigma}^+ c_{i,\sigma}\right) \\&-&
        |\Delta| \sum_i \left(e^{i\varphi_i} c_{i,\uparrow}^+
        c_{i,\downarrow}^+ + e^{-i\varphi_i} c_{i,\downarrow}
        c_{i,\uparrow}\right)
        ,
        \label{eq:H-BCS2}
\end{eqnarray}
where the summations run over all pairs $\langle i,j \rangle$ of
neighboring tight-binding sites in the kinetic energy given by
Eq.~(\ref{eq:H-BCS1}), and over all the tight-binding site labeled by
$i$ in the pairing term given by Eq.~(\ref{eq:H-BCS2}). The
superconducting phase variable is denoted by $\varphi_{i}$ in
Eq.~(\ref{eq:H-BCS2}) and the gap is denoted by $|\Delta|$. We assume
that no magnetic field penetrates in leads $S_a$, $S_b$, therefore
$\varphi_{i}$ is constant in each of them, with
$\varphi_{i}=\varphi_{a}$ in $S_a$ and $\varphi_{i}=\varphi_{b}$ in
$S_b$. We also assume that no magnetic flux penetrates in $S_c$, but
we choose to encode the Aharonov-Bohm flux $\Phi$ around the loop made
by $S_c$ through a pure gauge vector potential. As a result,
$\varphi_{i}$ varies inside $S_c$, and it takes values $\varphi_{c,1}$
and $\varphi_{c,2}$ at the two extremities of $S_c$, which are closest
to the dot.  Minimizing the condensate energy in the presence of the
Aharonov-Bohm vector potential in $S_c$ implies that
$\varphi_{c,2}-\varphi_{c,1}=\Phi$. Throughout {this
  paper, we use} the notation $\varphi_{c,1}=\varphi_{c}$, and
$\varphi_{c,2}=\varphi_{c}+\Phi$.

The coupling between the dot $x$ and each superconductor $S_p$ takes
the form of a usual tunneling Hamiltonian {with
  hopping amplitude $J_p$:}
\begin{equation}
  \label{eq:HJ}
  {\cal H}_{J_p} = J_p \sum_{\sigma} \int \frac{d^{3}\kbf}{(2\pi)^3}
  e^{-i s_p\omega_0t} c^+_{\sigma,p}(\kbf) d_{\sigma} +
  h.c.
 \end{equation}
Here $c^+_{\sigma,p}(\kbf)$ and $c_{\sigma,p}(\kbf)$ are creation and
annihilation operators for an electron on reservoir $p$ with momentum
$\kbf$ and spin $\sigma$ along the quantization axis. The
corresponding operators on the dot are denoted by $d^{+}_{\sigma}$ and
$d_{\sigma}$. We use the notation $\omega_0=eV/\hbar$.

The paper is focused on voltage biasing on the quartet line, according
to the experimental result of the Harvard group
\cite{Harvard-group-experiment}. This is why we use $V_j=s_j V$ for
the bias voltages. Specifically, the following values $s_a=1$,
{$s_b=-1$, $s_{c_1}=s_{c_2}=0$ are assigned to the}
parameters $s_j$, corresponding to voltage biasing at
$(V_a,V_b,V_{c,1},V_{c,2})=(V,-V,0,0)$.

We neglect quasiparticle tunneling through the loop from $S_{c,1}$ to
$S_{c,2}$, {\it i.e.} we assume that $S_{c,1}$ and $S_{c,2}$ are
solely coupled by the condensate of the grounded $S_c$. Since most of
the current is carried by Floquet resonances which are within the gap
of $S_c$, neglecting subgap quasiparticle processes through the loop
implies that the perimeter of the loop is large compared to {the BCS
  coherence length.}

\subsection{Infinite gap limit and gauge-invariant quartet phase}
\label{sec:infinite-gap}

This subsection presents the infinite gap limit and the
gauge-invariant phase variable. {Taking the limit of infinite gap was
  considered by many authors, see for instance
  Refs.~\onlinecite{Klees,Zazunov,Meng} to cite but a few. In our
  calculations, the Dyson equations produce a self-energy for the
  $2\times 2$ equilibrium quantum dot Green's functions, from which
  the following Hamiltonian is deduced in the Nambu representation:}
\begin{equation}
  \label{eq:H-infinite}
  {\cal H}_\infty=\left(
  \begin{array}{cc}
    0 & z\\
    \overline{z} & 0
  \end{array} \right)
  .
\end{equation}
Eq.~(\ref{eq:H-infinite}) implies two ABS at opposite energies $\pm
E_{ABS}$, with $E_{ABS}=|z|$.

The expression of $z$ is the following for a $(S_a,S_b,S_c)$ device
which is biased at the phases $(\varphi_a,\varphi_b,\varphi_c)$:
\begin{equation}
  \label{eq:z-3T}
  z_{3T}=\Gamma_a \exp\left(i\varphi_a\right) +\Gamma_b
  \exp\left(i\varphi_b\right) +\Gamma_c \exp\left(i\varphi_c\right)
  .
\end{equation}
The Josephson relations for three terminals $(S_a,S_b,S_c)$ biased at
$(V,-V,0)$ is given by Eqs.~(\ref{eq:k1})-(\ref{eq:k-def}).

The corresponding expression of $z_{4T}$ is the following with four
superconducting terminals $(S_a,S_b,S_{c,1},S_{c,2})$ which are
phase-biased at $(\varphi_a, \varphi_b, \varphi_{c,1},
\varphi_{c,2})$:
\begin{eqnarray}
  \label{eq:z-4T}
  z_{4T}&=&\Gamma_a \exp\left(i\varphi_a\right)
  +\Gamma_b \exp\left(i\varphi_b\right)\\
  &+&\Gamma_{c,1} \exp\left(i\varphi_{c,1}\right)
  +\Gamma_{c,2} \exp\left(i\varphi_{c,1}\right)
  ,
  \nonumber
\end{eqnarray}
and we used Eqs.~(\ref{eq:k1})-(\ref{eq:k-def}) for the superconducting
phases in the presence of voltage biasing.  We note that the
$(S_{c,1},S_{c,2})$ contacts can be gathered into a single $S_{c,eff}$
coupled by $\Gamma_{c,eff}$ to the dot, and with the phase
$\varphi_{c,eff}$:
\begin{equation}
  \label{eq:gamma-c-eff}
  \Gamma_{c,eff}\exp(i\varphi_{c,eff})=
  \Gamma_{c,1} \exp\left(i\varphi_{c,1}\right)
  +\Gamma_{c,2} \exp\left(i\varphi_{c,2}\right)
,
\end{equation}
with $\varphi_{c,eff}=\varphi_c+\alpha(\Phi)$, where $\alpha(\Phi)$
depends only on $\Phi$, {\it i.e.} it is independent on $\varphi_c$.
Then, all of the currents (which are gauge-invariant) depend on the
gauge-invariant quartet phase $\tilde{\varphi}_q$ which is expressed
as the following combination of the phase variables $\varphi_a$,
$\varphi_b$ and $\varphi_c$:
\begin{equation}
  \tilde{\varphi}_q=\varphi_q+\alpha(\Phi)
  \label{eq:varphiq-tilde}
  ,
\end{equation}
where the quartet phase is given by $\varphi_q=\varphi_a + \varphi_b -
2\varphi_c$.

\subsection{Quartet critical current}
\label{sec:quartet-current}

The expression of the quartet current is presented in this subsection.

The two-terminal dc-Josephson current is odd in the phase difference
\cite{Josephson}. In perturbation theory in the tunnel amplitudes, the
lowest-order quartet current is also odd in the superconducting
phases, and it is even in voltage. Generalizing to arbitrary values of
the contact transparencies, the quartet current
$I_q(eV/\Delta,\tilde{\varphi}_q/2\pi,\Phi/\Phi_0)$ is defined as the
component of
\begin{eqnarray}
  \label{eq:I-S-c}
I_{S_c}(eV/\Delta,\tilde{\varphi}_q/2\pi,\Phi/\Phi_0)&=&
I_{S_{c,1}}(eV/\Delta,\tilde{\varphi}_q/2\pi,\Phi/\Phi_0) \\&+&
I_{S_{c_2}}(eV/\Delta,\tilde{\varphi}_q/2\pi,\Phi/\Phi_0)
\nonumber
\end{eqnarray}
which is odd in $\tilde{\varphi}_q$ and in $\Phi$:
\begin{eqnarray}
    \label{eq:Iq-definition1}
  I_q(eV/\Delta,\tilde{\varphi}_q/2\pi,\Phi/\Phi_0) &=&
  I_{S_c}(eV/\Delta,\tilde{\varphi}_q/2\pi,\Phi/\Phi_0) \\&-&
  I_{S_c}(eV/\Delta,-\tilde{\varphi}_q/2\pi,-\Phi/\Phi_0) .
  \nonumber
\end{eqnarray}
Equivalently, $I_q(eV/\Delta,\tilde{\varphi}_q/2\pi,\Phi/\Phi_0)$ is
the component of Eq.~(\ref{eq:I-S-c}) which is even in voltage:
\begin{eqnarray}
  \label{eq:Iq-definition2}
I_q(eV/\Delta,\tilde{\varphi}_q/2\pi,\Phi/\Phi_0) &=&
I_{S_c}(eV/\Delta,\tilde{\varphi}_q/2\pi,\Phi/\Phi_0) \\&+&
I_{S_c}(-eV/\Delta,\tilde{\varphi}_q/2\pi,\Phi/\Phi_0) .
\nonumber
\end{eqnarray}
Eq.~(\ref{eq:Iq-definition2}) is used in the following numerical calculations.

The Harvard group experiment measures the critical current on the
quartet line for the device in figure~\ref{fig:device}, which we call
in short as ``the critical current'':
\begin{equation}
  \tilde{I}^*_{q,c}(eV/\Delta,\Phi/\Phi_0)=\mbox{Max}_{\tilde{\varphi}_q}
  I_q(eV/\Delta,\tilde{\varphi}_q/2\pi,\Phi/\Phi_0) ,
  \label{eq:Iqc-definition}
\end{equation}
where the quartet current $I_q(V,\tilde{\varphi}_q)$ is given by the
above Eqs.~(\ref{eq:Iq-definition1})-(\ref{eq:Iq-definition2}). Given
Eq.~(\ref{eq:varphiq-tilde}), taking the Max over $\tilde{\varphi}_q$
is equivalent to taking the Max over $\varphi_q$. This implies that
$\tilde{I}^*_{q,c}(eV/\Delta)$ is independent on $\alpha(\Phi)$. Thus,
it is only through $\Gamma_{c,eff}(\Phi)$ that
$\tilde{I}^*_{q,c}(eV/\Delta,\Phi)$ depends on $\Phi$.

\subsection{Parameters used in the numerical calculation}
\label{sec:parameters}

In this subsection, we present the parameters which are used in the
forthcoming numerical calculations.

Considering first a $(S_a,S_b,S_c)$ three-terminal Josephson junction,
the gap closes if the following condition on
$(\Gamma_a,\Gamma_b,\Gamma_c)$ is fulfilled \cite{theorie-quartets8}:
\begin{equation}
\Gamma_{c,eff}\exp(i\varphi_{c,eff}) =
  \frac{\left|\Gamma_a^2-\Gamma_b^2\right|}
       {\sqrt{\Gamma_a^2+\Gamma_b^2 -2 \Gamma_a \Gamma_b
           \cos\varphi_q}}
\label{eq:Gamma-c-eff-0}
  .
\end{equation}
Specializing to $\varphi_q=0$ leads to
\begin{eqnarray}
  \Gamma_{c,eff}&=&\Gamma_a+\Gamma_b\\
  \varphi_{c,eff}&=&0
  .
\end{eqnarray}

In the following numerical calculations, the four-dimensional
$(\Gamma_a,\Gamma_b,\Gamma_{c,1},\Gamma_{c,2})$ space of the coupling
constants between the dot and the superconducting leads is
scanned according to the following 1D subspace:
\begin{eqnarray}
  \label{eq:gamma-parameter-1}
  \frac{{\Gamma}_a}{\Delta}&=&0.4\\
  \label{eq:gamma-parameter-2}
  \frac{{\Gamma}_b}{\Delta}&=&0.2\\
  \label{eq:gamma-parameter-3}
  \frac{{\Gamma}_{c,1}}{\Delta}&=&\frac{1}{2}\left(0.3+\frac{\gamma}
  {\Delta}
  \right)\\
  \frac{{\Gamma}_{c,2}}{\Delta}&=&\frac{1}{2}\left(0.9
  +\frac{\gamma}{\Delta}\right)
  \label{eq:gamma-parameter-4}
  .
\end{eqnarray}
Eqs.~(\ref{eq:gamma-parameter-1})-(\ref{eq:gamma-parameter-4}) imply
\begin{equation}
  \Gamma_{c,1}+\Gamma_{c,2}-\Gamma_a-\Gamma_b = \gamma ,
\end{equation}
and the ABS gap closes at $\varphi_q=0$ if $\gamma/\Delta=0$. 

\subsection{Multilevel quantum dot}
\label{sec:multilevel-quantum-dot-model}

Now, we mention the multilevel quantum dot model containing $M$ energy
levels, used in section~\ref{sec:num2} in order to demonstrate
robustness of the inversion against multichannel effects.  This
multilevel quantum dot described in section~I of the Supplemental
Material is mapped onto an effective single-level quantum dot if a
specific condition of factorization is fulfilled.

\subsection{Inversion in the $V=0^+$ adiabatic limit}
\label{sec:adiabatic}
In this subsection, we mention section~II of the Supplemental
Material which provides a mechanism for the inversion in the $V=0^+$
adiabatic limit (still with biasing on the quartet line). It turns out
that inversion between $\Phi/\Phi_0=0$ and $\Phi/\Phi_0=1/2$
appears in the range of the $\Gamma$-parameters which
fulfills the conditions for convergence of perturbation theory in
$\Gamma_a$ and $\Gamma_b$ with respect to $\Gamma_{c,1}$ and
$\Gamma_{c,2}$, assumed to take much larger values. This predicted
inversion requires asymmetric couplings $\Gamma_{c,1}$ and
$\Gamma_{c,2}$.

However, this assumption on the couplings is not directly relevant to
{the situation where} the values of $\Gamma_{c,1}$ and $\Gamma_{c,2}$
{are more symmetric. Now, we select the parameters given by
  Eqs.~(\ref{eq:gamma-parameter-1})-(\ref{eq:gamma-parameter-4}) which
  produce ``absence of inversion in the $V=0^+$ adiabatic limit'' and
  investigate a mechanism for emergence of inversion at finite bias
  voltage $V$.}

\begin{figure*}[htb]
  \includegraphics[width=\columnwidth]{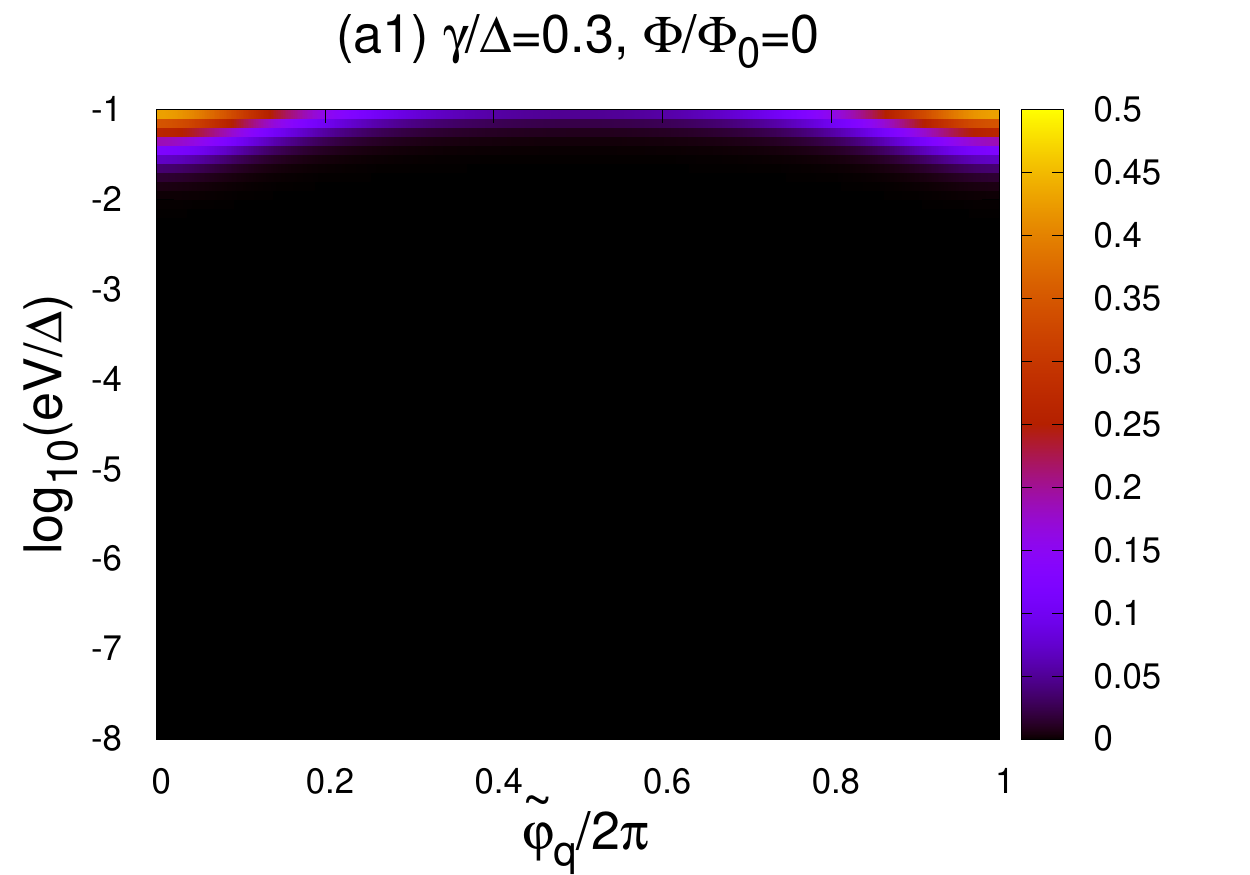}\includegraphics[width=\columnwidth]{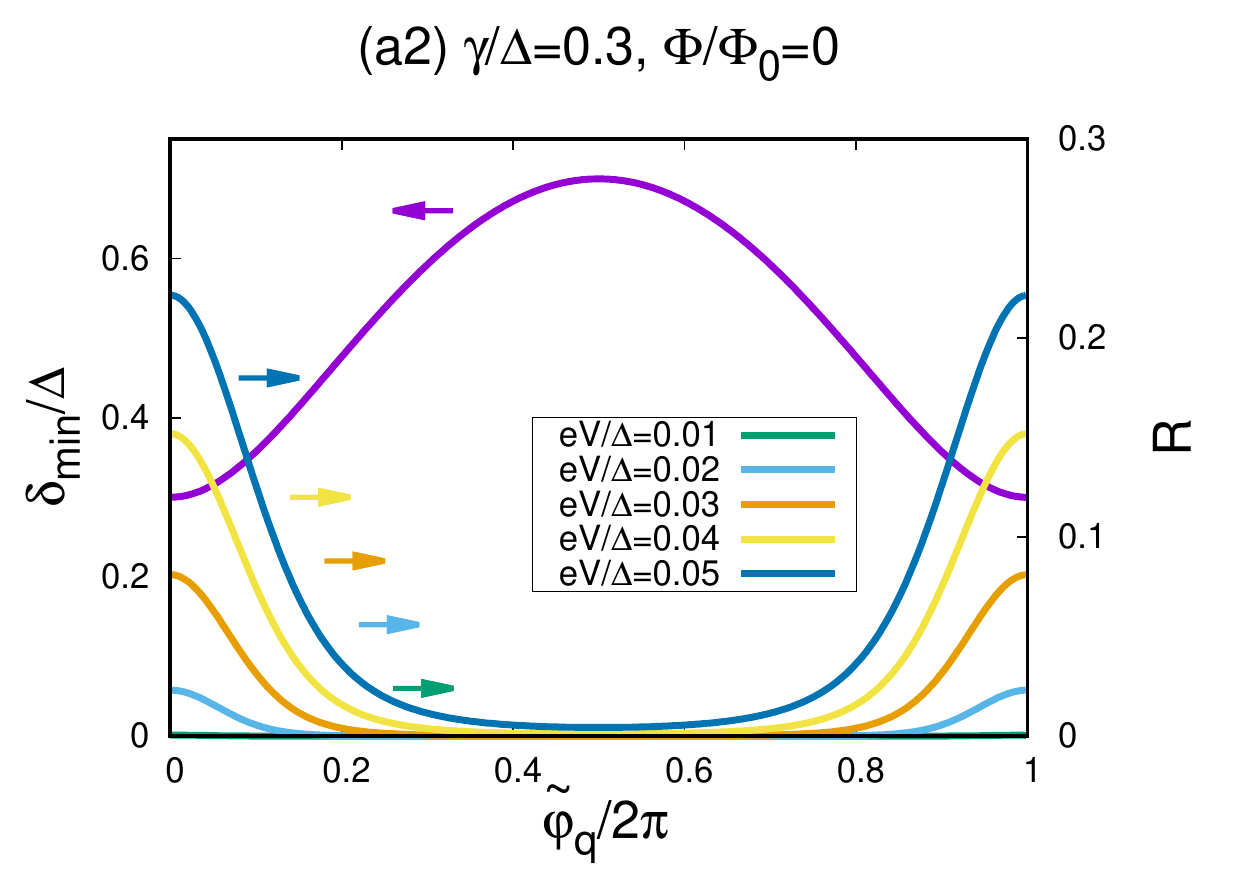}

  \includegraphics[width=\columnwidth]{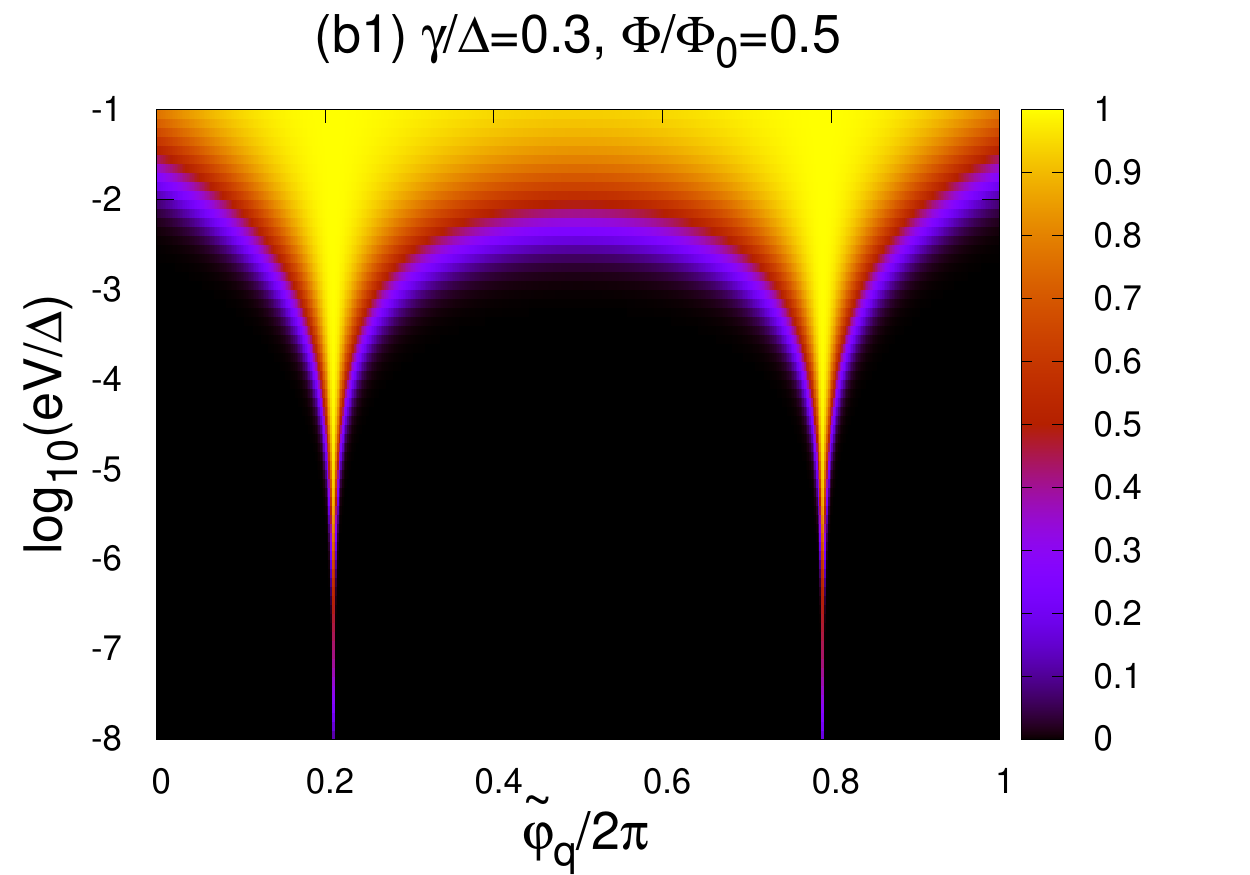}\includegraphics[width=\columnwidth]{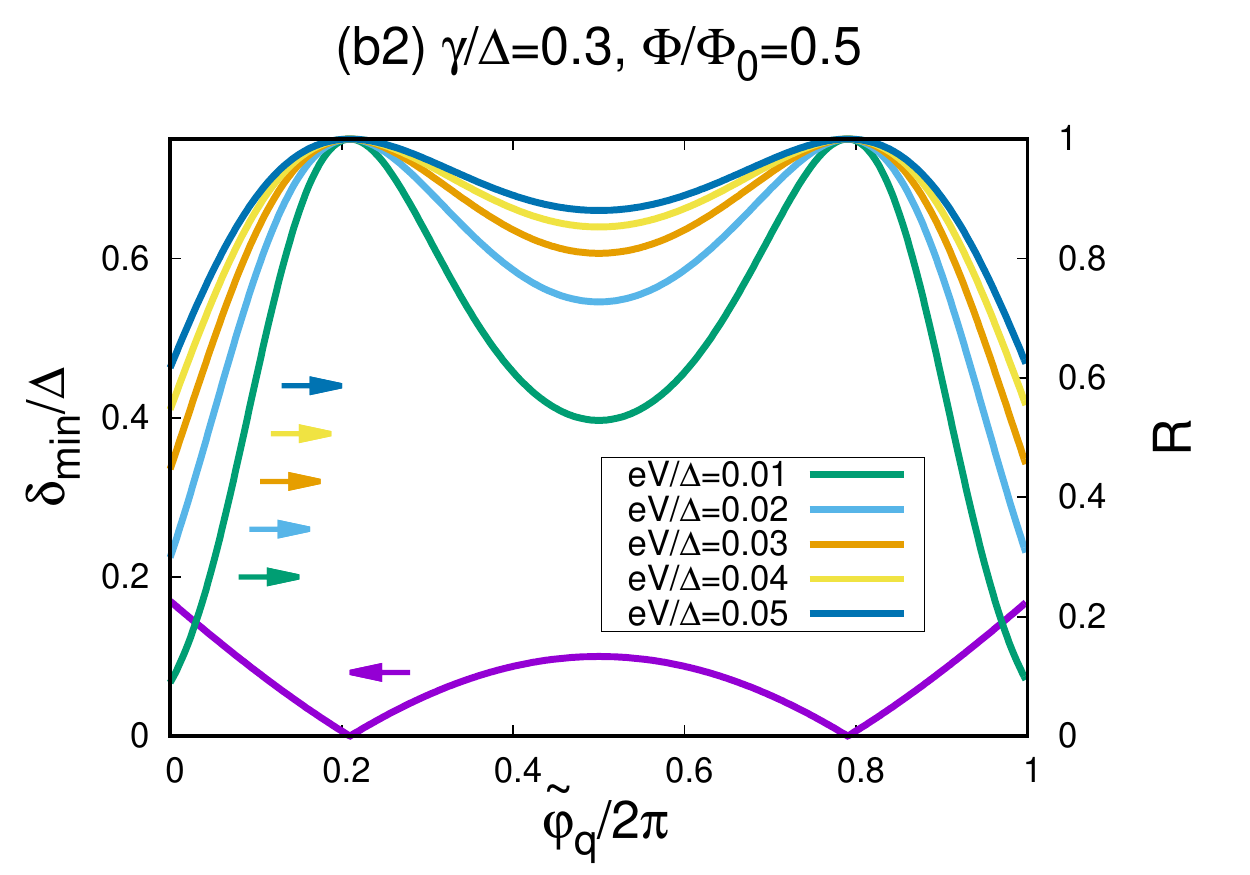}

  \includegraphics[width=\columnwidth]{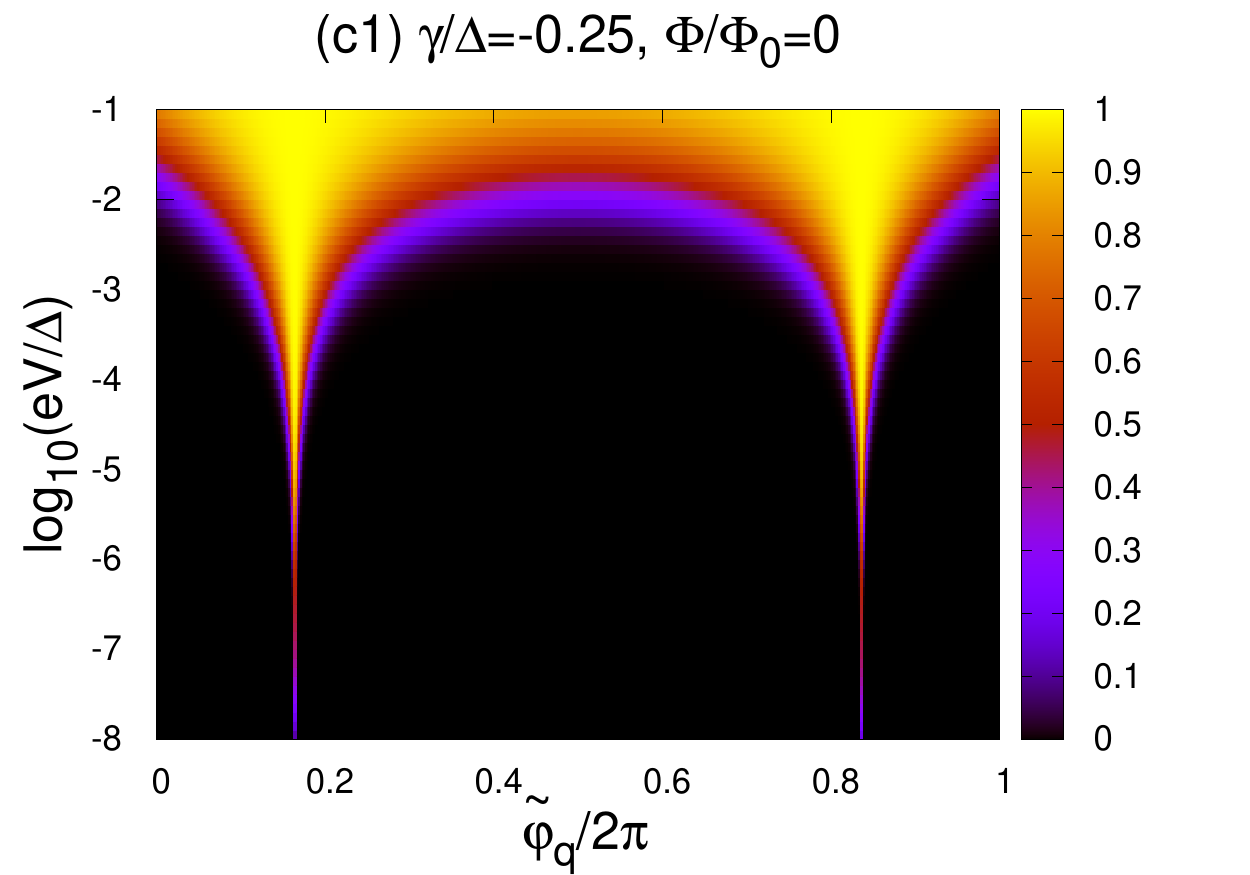}\includegraphics[width=\columnwidth]{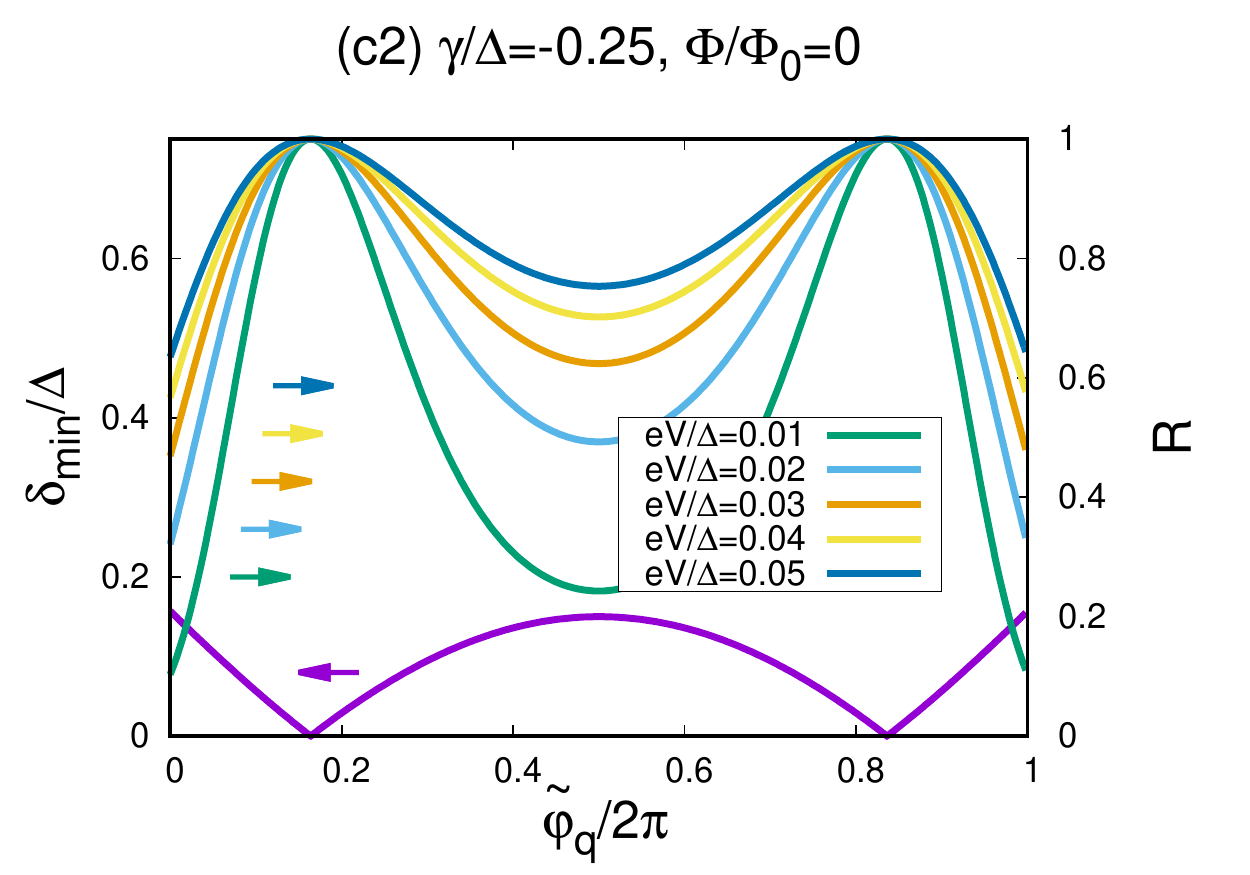}
  \caption{{\it The Landau-Zener tunneling rates:} The figure shows
    the rate ${\cal R}$ of Landau-Zener tunneling for
    $\gamma/\Delta=0.3$ and $\Phi/\Phi_0=0$ (panels a1 and a2),
    $\gamma/\Delta=0.3$ and $\Phi/\Phi_0=1/2$ (panels b1 and b2) and
    $\gamma/\Delta=-0.25$ and $\Phi/\Phi_0=0$ (panels c1 and
    c2). Panels a1, b1 and c1 show colorplots of ${\cal R}$ as a
    function of $\tilde{\varphi}_q/2\pi$ (on $x$-axis) and
    $\log_{10}(eV/\Delta)$ (on $y$-axis). Panels a2, b2 and c2 show
    $\delta_{min}/\Delta$ (magenta lines) as a function of
    $\tilde{\varphi}_q/2\pi$, and ${\cal R}(\tilde{\varphi}_q/2\pi)$
    evaluated for $eV/\Delta=0.01,\,0.02,\,0.03,\,0.04,\,0.05$.
    \label{fig:R}
  }
\end{figure*}

\section{Landau-Zener tunneling rate}
\label{sec:R}

{This section provides the calculations of the Landau-Zener tunneling
  rate ${\cal R}$. Evaluating of ${\cal R}$ is used as a
  ``calibration'' to select a few values of the device parameters
  representative of ``weak'' and ``strong'' Landau-Zener
  tunneling. Next, the selected values of the $\Gamma$s [see
    Eqs.~(\ref{eq:gamma-parameter-1})-(\ref{eq:gamma-parameter-4})]
  and $\Phi/\Phi_0$ will be implemented to obtain the Floquet spectra
  and the quartet critical current in the next sections~\ref{sec:num1}
  and~\ref{sec:num2}.}

Subsection~\ref{sec:R-analytical} presents the analytical calculations
of ${\cal R}$. Section~\ref{sec:R-numerics} shows a numerical
illustration with the parameters of the forthcoming
sections~\ref{sec:num1} and~\ref{sec:num2}.

\subsection{Analytical results}
\label{sec:R-analytical}

In this subsection, we present an analytical theory of an indicator
for the strength of quantum fluctuations in the quartet current: the
rate ${\cal R}$ of Landau-Zener tunneling between the two ABS
manifolds.

It was shown in section~\ref{sec:model-and-Hamiltonian} that the
four-terminal device on figure~\ref{fig:device} can be mapped onto
three terminals with suitable coupling $\Gamma_{c,eff}$ between the
dot and the grounded lead $S_{c,eff}$ [see
  Eq.~(\ref{eq:gamma-c-eff})]. Thus, the Landau-Zener tunneling rate
${\cal R}$ is now evaluated for a three-terminal device, without loss
of generality with respect to four terminals. We use the
notation $k$ {for the fast combination of the
  superconducting phases, see
  Eqs.~(\ref{eq:k1})-(\ref{eq:k4}). Eq.~(\ref{eq:z-3T})} leads to the
following expression for the ABS energies:
\begin{equation}
  \label{eq:E-ABS-3T}
  E_{ABS,3T}=\left|\Gamma_{3T}\right|=\left|\Gamma_{a,3T} e^{i(\varphi_a+k)}
  +\Gamma_{b,3T} e^{i(\varphi_b-k)}
  +\Gamma_{c,3T} e^{i\varphi_c}\right|
  .
\end{equation}
We first evaluate the value $k_*$ of $k$ which minimizes $E_{ABS,3T}$
in Eq.~(\ref{eq:E-ABS-3T}). The corresponding energy at the minimum is
denoted by $\delta_{min}$:
\begin{equation}
  \label{eq:delta-min}
  \delta_{min}=\mbox{Inf}_k \left[E_{ABS,3T}(k)\right]
  ,
\end{equation}
which depends on all junction parameters. Eq.~(\ref{eq:delta-min}) can
be called as ``{\it the Andreev gap}'' if the ABS spectrum is plotted
as a function of the fast variable $k$. We have shown previously
\cite{theorie-quartets8} that a single or two local minima can occur
in the variations of $E_{ABS}$ with $k$, depending on the values of
the device parameters. As a simplifying assumption, the Landau-Zener
processes are considered to be dominated by the global minimum in the
presence of two local minima. In a second step, $E_{ABS}$ given by
Eq.~(\ref{eq:E-ABS-3T}) is expanded to second order in the vicinity of
$k_*$:
\begin{equation}
  E_{ABS}^2= \delta_{min}^2 +
  \tilde{\Gamma}_0^2 \left(k-k_*\right)^2 + {\cal O}\left[
    \left(k-k_*\right)^3\right]
  ,
\end{equation}
where the coefficient $\tilde{\Gamma}_0$ is the following:
\begin{eqnarray}
  \tilde{\Gamma}_0^2&=&-4\Gamma_{a,3T}\Gamma_{b,3T} \cos\left(
  2k_*-\varphi_a+\varphi_b\right)\\
  \nonumber
  &-&\Gamma_{a,3T}\Gamma_{c,3T} \cos\left(k_*-\varphi_a\right)
  -\Gamma_{b,3T}\Gamma_{c,3T} \cos\left(k_*+\varphi_b\right)
  .
\end{eqnarray}
The rate ${\cal R}$ of Landau-Zener tunneling can be
approximated as the following:
\begin{equation}
  \label{eq:expression-of-R}
  {\cal R}=\exp\left(-\frac{\pi \delta_{min}^2}{4eV\tilde{\Gamma}_0}\right)
  .
\end{equation}
Eq.~(\ref{eq:expression-of-R}) appeared previously in the literature,
see for instance Eq.~(20) in a review article on
Landau-Zener-St\"uckelberg interferometry \cite{Shevchenko}.

\subsection{Numerical results}
\label{sec:R-numerics}

In this subsection, we present figure~\ref{fig:R} showing numerical
illustration for the rate ${\cal R}$ of Landau-Zener tunneling [see
  Eq.~(\ref{eq:expression-of-R})].

Figures~\ref{fig:R}~a1, b1 and c1 show colorplots of ${\cal R}$ in the
plane of the reduced parameters
$\left(\tilde{\varphi}_q/2\pi,\log_{10}(eV/\Delta)\right)$. The
following parameters are used: $\gamma/\Delta=0.3$, $\Phi/\Phi_0=0$
(panel a1), $\gamma/\Delta=0.3$, $\Phi/\Phi_0=1/2$ (panel b1), and
$\gamma/\Delta=-0.25$, $\Phi/\Phi_0=0$ (panel c1).  The yellow
colorcode on figures~\ref{fig:R}~a1, b1 and c1 corresponds to strong
Landau-Zener tunneling with ${\cal R}\simeq 1$. The black colorcode
corresponds to the adiabatic limit with negligibly small Landau-Zener
tunneling ${\cal R}\simeq 0$.

Figures~\ref{fig:R}~a2, b2 and c2 represent the ``Andreev gap''
$\delta_{min}$ as a function of the gauge-invariant quartet phase
$\tilde{\varphi}_q$ for the same parameters as figures~\ref{fig:R}~a1,
b1 and c1 (see above). In addition, figures~\ref{fig:R}~a2, b2 and c2
show the variations of ${\cal R}$ with $\tilde{\varphi}_q/2\pi$, for
the following values of voltage:
$eV/\Delta=0.01,\,0.02,\,0.03,\,0.04,\,0.05$.  These reduced voltage
values $eV/\Delta$ are close to those of the forthcoming
sections~\ref{sec:num1} and~\ref{sec:num2}.

Considering now interpretation of figure~\ref{fig:R}, the rate ${\cal
  R}$ of Landau-Zener tunneling given by
Eq.~(\ref{eq:expression-of-R}) has exponential variations with all of
the following parameters: the reduced voltage $eV/\Delta$, the reduced
flux-$\Phi/\Phi_0$, the gauge-invariant quartet phase
$\tilde{\varphi}_q$, and the parameter $\gamma/\Delta$ used to
parameterize the coupling between the dot and the superconducting
leads [see
  Eqs.~(\ref{eq:gamma-parameter-1})-(\ref{eq:gamma-parameter-4})]. The
exponential dependence is compatible with the narrow cross-over along
the $y$-voltage axis on figure~\ref{fig:R}, between the low-voltage
adiabatic and the higher-voltage antiadiabatic regimes of the black
and yellow color-codes respectively.

Figures~\ref{fig:R}~a1, b1, c1 correlate with the gauge-invariant
quartet phase $\tilde{\varphi}_q/2\pi$-sensitivity of the Andreev gap
$\delta_{min}$ on figures~\ref{fig:R}~a2, b2 and c2
respectively. Namely, closing the Andreev gap $\delta_{min}$ at
$\tilde{\varphi}_q/2\pi$ around $\tilde{\varphi}_q/2\pi\simeq
0.2,\,0.8$ (magenta line on panels b2, c2) results in strong
nonadiabaticity.  The Andreev gap $\delta_{min}$ does not close at any
value of $\tilde{\varphi}_q/2\pi$ for weak Landau-Zener (see the
magenta line on figure~\ref{fig:R} a1). Panel a1 shows ${\cal R}\simeq
0$ in most of the considered voltage range $-8\le
\log_{10}(eV/\Delta)\le -1$ while yellow-colored regions with ${\cal
  R}\simeq 1$ clearly develop on panels b1, c1.

\begin{figure*}[htb]
    \includegraphics[width=\textwidth]{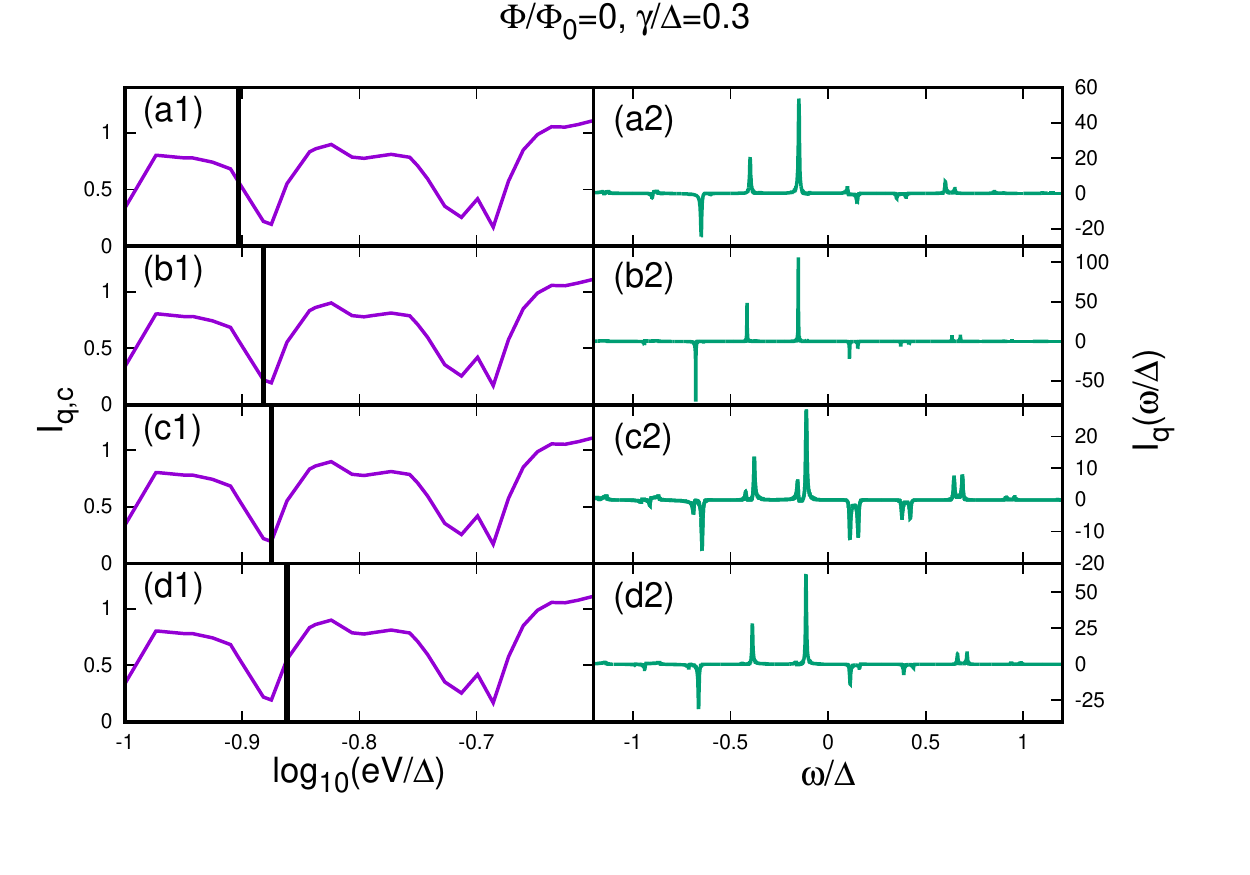}
    \caption{{\it The spectral current:} Panels a1-d1 show the quartet
      critical current $I_{q,c}$ as a function of the log of the
      reduced voltage $\log_{10}(eV/\Delta)$. The vertical bars on
      panels a1-d1 indicate the values of the voltages which are
      selected on panels a2-d2. The latter show the spectral current
      at these $eV/\Delta$-values as a function of reduced energy
      $\omega/\Delta$.  The figure corresponds to $\Phi/\Phi_0=0$ and
      $\gamma/\Delta=0.3$, {\it i.e.} to weak Landau-Zener tunneling.
    \label{fig:spectral_currents}
  }
\end{figure*}

\begin{figure*}[htb]
  \includegraphics[width=\textwidth]{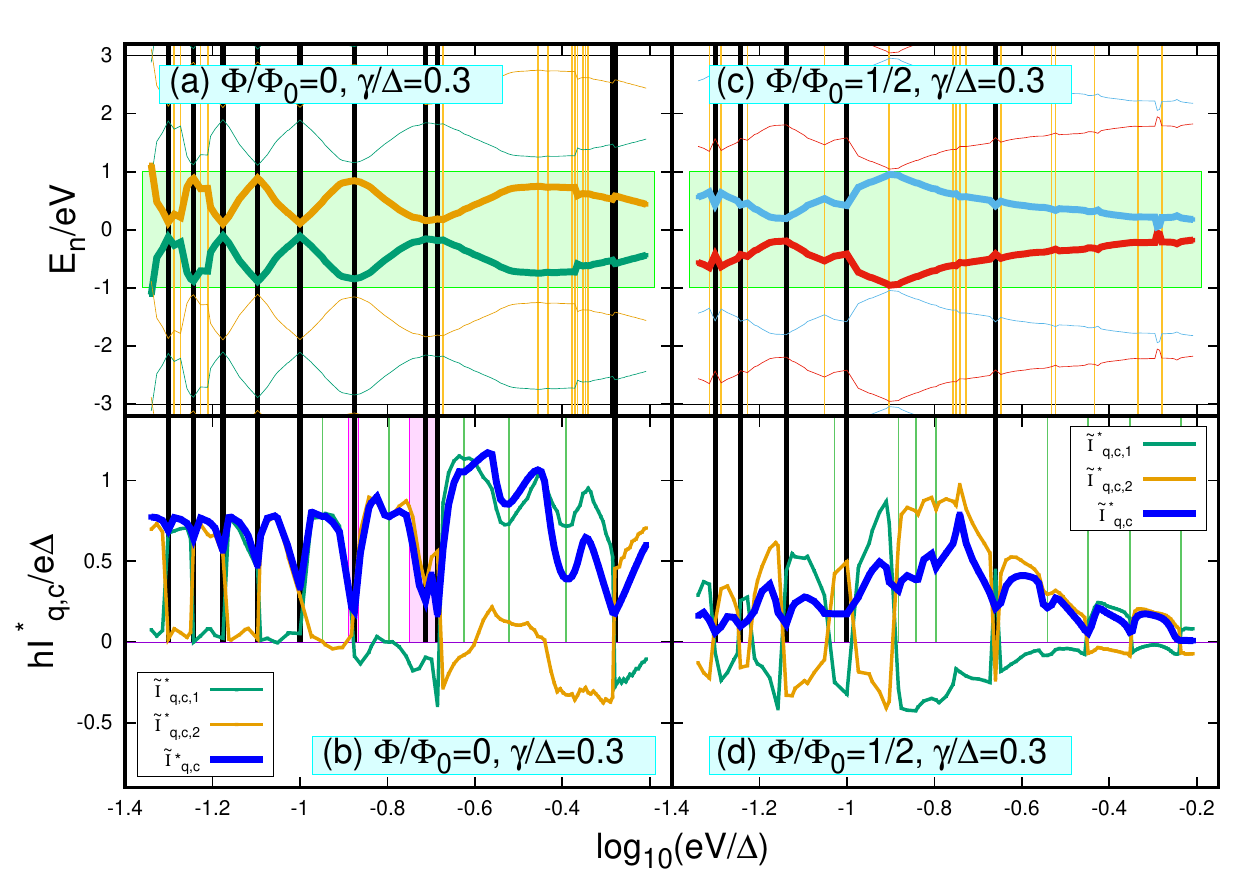}
  \caption{{\it Correspondence between the Floquet spectra and the
      quartet critical current:} The figure shows the Floquet spectra
    (panels a and c) and the critical quartet current $I_{q,c}$
    (panels b and d) as a function of $\log_{10}(eV/\Delta)$ on
    $x$-axis, for $\gamma/\Delta=0.3$ and $\Phi/\Phi_0=0$ (panels a
    and b, being representative of ``weak Landau-Zener'') and for
    $\gamma/\Delta=0.3$ and $\Phi/\Phi_0=1/2$ (panels c and d being
    representative of ``strong Landau-Zener''). Panels b and d also
    show the currents $I_{q,c,1}$ and $I_{q,c,2}$ carried by each
    Floquet state [see Eqs.~(\ref{eq:I1}) and~(\ref{eq:I2})]. The
    vertical bars show the extrema in the Floquet spectra (on panels a
    and b) and the minima in $I_{q,c}$ (on panels b and d), with the
    following colorcode:~(i) ``Black vertical bars'' are used for
    coincidence between the extrema in the Floquet spectra and the
    minima in $I_{q,c}$,~(ii) ``Orange vertical bars'' on panels a and
    c are used for the extrema in the Floquet spectra which have no
    counterpart as a minimum in $I_{q,c}$,~(iii) ``Green vertical
    bars'' on panels b and d are used for the minima in $I_{q,c}$
    which have no counterpart as an extremum in the Floquet spectrum.
    \label{fig:low_voltage_window1}
    }
\end{figure*}

To summarize, we calculated the variations of the Landau-Zener
tunneling rate ${\cal R}$ for the three sets of parameters which will
be used in the next sections~\ref{sec:num1} and~\ref{sec:num2}. One of
those is representative of ``weak Landau-Zener tunneling''
characterized by a finite Andreev gap in the entire
$\tilde{\varphi}_q/2\pi$-parameter range, {\it i.e.}
$\gamma/\Delta=0.3$ and $\Phi/\Phi_0=0$ on figures~\ref{fig:R}~a1-a2.
The two others correspond to ``strong Landau-Zener tunneling''
characterized by closing the ``Andreev gap'' at specific values of
$\tilde{\varphi}_q/2\pi$, {\it i.e.} $\gamma/\Delta=0.3$ and
$\Phi/\Phi_0=1/2$ on figures~\ref{fig:R}~b1-b2 and
$\gamma/\Delta=-0.25$ and $\Phi/\Phi_0=0$ on
figures~\ref{fig:R}~c1-c2.

\section{Inversion at finite bias voltage $V\ne 0$}
\label{sec:num1}

Now, we present the main results and discuss how Landau-Zener
tunneling can produce inversion between $\Phi/\Phi_0=0$ and
$\Phi/\Phi_0=1/2$, {\it i.e.}
$I_{q,c}(eV/\Delta,0)<I_{q,c}(eV/\Delta,1/2)$.

The algorithms are mentioned in section~\ref{sec:algorithms}. The
quartet critical current is defined in section~\ref{sec:definition}.
Section~\ref{sec:numerical-experiments} presents the numerical
data which are next discussed physically in
section~\ref{sec:the-physical-picture}. A summary is presented in
section~\ref{sec:summary}.

\subsection{Algorithms}
\label{sec:algorithms}

{ The code is based on Ref.~\onlinecite{Cuevas}, and it was developed over
  the last years to address Floquet theory in multiterminal quantum
  dot Josephson junctions, in connection with the dc-quartet current,
  zero- and finite-frequency noise
  \cite{theorie-quartets5,theorie-quartets6,theorie-quartets7,theorie-quartets8,Heiblum}.}
The principle of the code is summarized in the Appendix of
Ref.~\onlinecite{theorie-quartets5}.

{In short, the dc-current} $I_{S_c}=I_{S_{c,1}}+I_{S_{c,2}}$ entering
the grounded $S_c$ is evaluated from integral over the energy $\omega$
of the spectral current ${\cal I}_{S_c}(\omega)$:
\begin{equation}
  \label{eq:ISc-spectral}
  I_{S_c}=\int {\cal I}_{S_c}(\omega) d\omega
  .
\end{equation}
The spectral quartet current ${\cal I}_{S_c}(\omega)= {\cal
  I}_{S_{c,1}}(\omega)+ {\cal I}_{S_{c,2}}(\omega)$ transmitted into
$S_{c,1}$ and $S_{c,2}$ is calculated from the Keldysh Green's
function.  Adaptative algorithm is used to integrate over $\omega$,
and matrix multiplications are optimized with sparse matrix
algorithms.

The spectral current shows sharp peaks at the energies $\omega=E_n$ of
the Floquet levels
\cite{theorie-quartets6,theorie-quartets7,theorie-quartets8,Domanski2}.
Figures~\ref{fig:spectral_currents}~a2-d2 show how the peaks in the
quartet spectral current ${\cal I}_{q}(\omega)$ deduced from ${\cal
  I}_{S_c}(\omega)$ evolve as the reduced voltage $eV/\Delta$
indicated on figures~\ref{fig:spectral_currents}~a1-d1 is scanned
through a dip in $I_{q,c}(eV/\Delta,\Phi/\Phi_0)$. Further comments
about this figure are presented in section~\ref{sec:pi-shift}, in
connection with populations of the Floquet states.

\subsection{Definition of the quartet critical current as a function
  of voltage}
\label{sec:definition}

Now, we define a central quantity: the quartet critical current as a
function of reduced voltage $eV/\Delta$.

The value of the gauge-invariant quartet phase $\tilde{\varphi}_q$ is
calculated in such a way as to maximize the current
$I_{S_c}=I_{S_{c,1}}+I_{S_{c,2}}$ transmitted into the grounded loop
$S_c$ at the contacts points $S_{c,1}$ and $S_{c,2}$, as a function of
the gauge invariant quartet phase $\tilde{\varphi}_q$. The value of
$\tilde{\varphi}_q$ which maximizes the current is denoted by
$\tilde{\varphi}_q^*$. In the spirit of Eq.~(\ref{eq:Iqc-definition}),
the value of the current at the maximum is denoted by
\begin{eqnarray}
\label{eq:quartet-critical-current}
&&  \tilde{I}_{q,c}^*(eV/\Delta,\Phi/\Phi_0) =
  \tilde{I}_{S_c}(eV/\Delta,\tilde{\varphi}_q^*/2\pi,\Phi/\Phi_0)\\
  \nonumber
&=&  I_{S_{c,1}}(eV/\Delta,\tilde{\varphi}_q^*/2\pi,\Phi/\Phi_0) +
  I_{S_{c,2}}(eV/\Delta,\tilde{\varphi}_q^*/2\pi,\Phi/\Phi_0)\\
\nonumber
  &=&
  \mbox{Max}_{\tilde{\varphi}_q}
  \left[I_{S_{c,1}}(eV/\Delta,\tilde{\varphi}_q/2\pi,\Phi/\Phi_0)\right.\\
    \nonumber
   && + \left.
    I_{S_{c,2}}(eV/\Delta,\tilde{\varphi}_q/2\pi,\Phi/\Phi_0)\right] .
\end{eqnarray}
The quantity $\tilde{I}_{q,c}^*(eV/\Delta,\Phi/\Phi_0)$ is called in short
as ``the critical current''. 

Now, we present the currents
$\tilde{I}^*_{q,c,1}(eV/\Delta,\Phi/\Phi_0)$ and
$\tilde{I}^*_{q,c,2}(eV/\Delta,\Phi/\Phi_0)$ carried by each Floquet
state.

Specifically, the spectral current $\tilde{I}(\omega)$ is ``folded''
into the first Brillouin zone $[0,2eV]$
\begin{equation}
  \label{eq:Itilde-folded}
\tilde{I}_{folded}(\tilde{\omega})=\sum_n \tilde{I}(\tilde{\omega}+2neV)
,
\end{equation}
where $0 < \tilde{\omega} < 2eV$ in Eq.~(\ref{eq:Itilde-folded}). The
currents $\tilde{I}_1$ and $\tilde{I}_2$ carried by each Floquet state
are the contributions of the $0 < \tilde{\omega} < eV$ and the $eV <
\tilde{\omega} < 2eV$ spectral windows:
\begin{eqnarray}
  \label{eq:I1}
\tilde{I}_1&=&\int_0^{eV} \tilde{I}_{folded}(\tilde{\omega}) d\tilde{\omega}\\
\tilde{I}_2&=&\int_{eV}^{2eV} \tilde{I}_{folded}(\tilde{\omega}) d\tilde{\omega}
.
\label{eq:I2}
\end{eqnarray}
The values of $\tilde{I}_1$ and $\tilde{I}_2$ at $\tilde{\varphi}_q =
\tilde{\varphi}_q^*$ are denoted by $\tilde{I}_{q,c,1}^*$ and
$\tilde{I}_{q,c,2}^*$ respectively. The contributions
$\tilde{I}_{q,c,1}^*$ and $\tilde{I}_{q,c,2}^*$ of the Floquet states
$1$ and $2$ are calculated solely from maximizing the total current
$\tilde{I} = \tilde{I}_{1} + \tilde{I}_{2}$ with respect to
$\tilde{\varphi}_q$, not from separately maximizing $\tilde{I}_1$ and
$\tilde{I}_2$.

Concerning the choice of the parameters, this section~\ref{sec:num1}
discusses solely ``weak Landau-Zener tunneling'' for
$\gamma/\Delta=0.3$ and $\Phi/\Phi_0=0$ (corresponding to
figures~\ref{fig:R}~a1, a2 in the preceding section~\ref{sec:R}). The
discussion of strong Landau-Zener tunneling (such as for
$\gamma/\Delta=0.3$ and $\Phi/\Phi_0=1/2$) is postponed for
section~\ref{sec:num2}.

\subsection{Presentation of the numerical results}
\label{sec:numerical-experiments}
Now, we show our numerical data in themselves, and we postpone the
physical discussion to section~\ref{sec:the-physical-picture} in
the continuation of the previous section~\ref{sec:summary-of-the-paper}.

The Floquet spectra are presented in
section~\ref{sec:Floquet-spectra}. The critical current is presented
in section~\ref{sec:critical-current}. The connection between the
Floquet spectra and the critical current is presented in
section~\ref{sec:connection-bars}.

\subsubsection{Numerical results for the Floquet spectra}
\label{sec:Floquet-spectra}

{The Floquet spectra were introduced in
  section~\ref{sec:summary-of-the-paper}, starting with the quantum
  Landau-Zener tunneling on top of the classical $V=0^+$ adiabatic
  limit. Now, we present the actual numerical data, focusing on
  evidence for avoided crossings.}

Figure~\ref{fig:low_voltage_window1} shows comparison between~(i)
The Floquet energies $E_n$ as a function of $\log_{10}(eV/\Delta)$,
and~(ii) The critical current $I_{q,c}$. The values $\Phi/\Phi_0=0$
and $\Phi/\Phi_0=1/2$ of the reduced flux are used on panels a-b and
c-d respectively, and the contact transparencies are such that
$\gamma/\Delta=0.3$ in
Eqs.~(\ref{eq:gamma-parameter-1})-(\ref{eq:gamma-parameter-4}), {\it
  i.e.} they are relevant to weak Landau-Zener tunneling according to
section~\ref{sec:summary-of-the-paper}.

Figure~\ref{fig:low_voltage_window1}a shows the normalized
Floquet energies $E_n/eV$ as a function of the reduced voltage
$eV/\Delta$. The dynamics
is periodic in time with period $\hbar/2eV$ and the Floquet spectrum
is periodic in energy with period $2eV$.  The shaded green region on
panels a and c show the ``first Brillouin zone'' $-1 < E_n/eV <
1$. The other Floquet levels are obtained by translation along the
$y$-axis of energy according to $\{E_{-1}+2peV,\,E_1+2qeV\}$ with $p$
and $q$ two integers, where $-eV<E_{-1}<0$ and $0<E_1<eV$.

Following the previous section~\ref{sec:summary-of-the-paper}, we note
that, as on figure~\ref{fig:supplementary-1}~b, the quantum mechanical
Landau-Zener tunneling opens gaps in the Floquet spectrum in
figure~\ref{fig:low_voltage_window1}a, instead of the classically
nonavoided level crossings at $\{e V_{cross,n}\}$ on
figure~\ref{fig:supplementary-1}~a.

\subsubsection{Numerical results for the critical current}
\label{sec:critical-current}

Now, we comment on the critical current
$\tilde{I}^*_{q,c}(eV/\Delta,\Phi/\Phi_0)$ defined by
Eq.~(\ref{eq:quartet-critical-current}) in the previous
section~\ref{sec:definition}.  The variations of
$\tilde{I}^*_{q,c}(eV/\Delta,\Phi/\Phi_0)$ with $\log_{10}(eV/\Delta)$
are shown by the blue lines in
figure~\ref{fig:low_voltage_window1}b. Figure~\ref{fig:low_voltage_window1}b
reveals a regular sequence of ``dips'' in the reduced
voltage-$eV/\Delta$ dependence of
$\tilde{I}^*_{q,c}(eV/\Delta,\Phi/\Phi_0)$, which is in a qualitative
agreement with the mechanism discussed in the preceding
section~\ref{sec:summary-of-the-paper}, see
figures~\ref{fig:supplementary-2}~a-b.  The discussion of the
contributions $\tilde{I}^*_{q,c,1}(eV/\Delta,\Phi/\Phi_0)$ and
$\tilde{I}^*_{q,c,2}(eV/\Delta,\Phi/\Phi_0)$ of each Floquet state
(green and orange lines on figure~\ref{fig:low_voltage_window1}b) is
postponed for section~\ref{sec:the-physical-picture} below.

\subsubsection{Numerical evidence for a
  connection between the Floquet spectra and the
  current}
\label{sec:connection-bars}

Now, we present a connection between the Floquet spectra and the
quartet current according to figures~\ref{fig:supplementary-2}~a-b in
the preceding section~\ref{sec:summary-of-the-paper}, {\it i.e.} we
discuss the vertical bars in
figures~\ref{fig:low_voltage_window1}~a-b:

(i) The extrema in the Floquet spectra are shown by the vertical bars
on figure~\ref{fig:low_voltage_window1}a. They are such that $\partial
E_n(V_{Fl,\lambda})/\partial V=0$ (where the integer $\lambda$ labels
the extrema).

(ii) The minima in $\tilde{I}^*_{q,c,\mu}(V)$ are shown by the
vertical bars on figure~\ref{fig:low_voltage_window1}b. They are such
that $\partial \tilde{I}^*_{q,c}(V_{q,c,\mu})/\partial V=0$ and
$\partial^2 \tilde{I}^*_{q,c}(V_{q,c,\mu})/\partial V^2>0$ (where the
integer $\mu$ labels the minima).

The following colorcode is used for these vertical bars:

(i) The black vertical bars on
figure~\ref{fig:low_voltage_window1}~a-b show the voltage-$V$ values
such that $V_{Fl,\lambda}\simeq V_{q,c,\mu}$ are coinciding within a
small tolerance.

(ii) The thinner vertical orange bars on
figure~\ref{fig:low_voltage_window1}a show the values of
$V_{Fl,\lambda}$ which are noncoinciding with any of the
$\{V_{q,c,\mu}\}$.

(iii) The thinner vertical magenta bars on
figure~\ref{fig:low_voltage_window1}b show the values of $V_{q,c,\mu}$
which are noncoinciding with any of the $\{V_{Fl,\lambda}\}$.

\subsection{The physical picture}
\label{sec:the-physical-picture}

Subsection~\ref{sec:numerical-experiments} presents the numerical data
for $\gamma/\Delta=0.3$, {\it i.e.} with small Landau-Zener tunneling
rate. Now, we discuss physically the data shown in the preceding
section~\ref{sec:connection-bars}, in connection with the above
section~\ref{sec:summary-of-the-paper}.

{In short, three regimes} are obtained upon increasing
voltage $V$ from the $V=0^+$ adiabatic limit, {\it i.e.} upon
increasing the strength of Landau-Zener tunneling:

(i) At low voltage, Landau-Zener tunneling implies {\it hybridization}
between the Floquet states at the avoided crossings in the Floquet
spectrum (see section~\ref{sec:hybridization}).

(ii) Increasing voltage has the effect of enhancing Landau-Zener
tunneling and populating both Floquet states.

(iii) At higher voltage, the nontrivial populations of the Floquet
states produce {\it $\pi$-shifted current-phase relations} (see
section~\ref{sec:pi-shift}).

\subsubsection{Hybridization between the two Floquet states
  at very low voltage}
\label{sec:hybridization}

{\it Connection between the Floquet spectra and the quartet current: }
{We discuss now} the coincidence $V_{Fl,\lambda}=V_{q,c,\mu}$ reported
in the preceding section~\ref{sec:connection-bars}. {The notation
  $V_{Fl,\lambda}$ is used for the values of the voltage corresponding
  to the extrema in the Floquet spectrum ({\it i.e.} the voltages of
  the avoided crossings), and $V_{q,c,\mu}$ denote the voltage-values
  of the minima in the quartet critical current $I_{q,c}(V)$.  The
  correspondence between the voltage-$V$ dependence of the Floquet
  spectrum and the quartet critical current $I_{q,c}(V)$ is
  interpreted as a common physical mechanism of Landau-Zener
  tunneling, see the above section~\ref{sec:summary-of-the-paper}:}

(i) Landau-Zener tunneling produces quantum mechanical coupling
between the two Floquet states. The two ABS at opposite energies
contribute for opposite values to the currents
$I_{S_{c,1}}(eV/\Delta, \tilde{\varphi}_q, \Phi/\Phi_0)$ and
$I_{S_{c,2}}(eV/\Delta, \tilde{\varphi}_q, \Phi/\Phi_0)$ at the
$S_{c,1}$ and $S_{c,2}$ contacts. Thus, Landau-Zener tunneling reduces
the critical current $I_{q,c}$ by quantum mechanically coupling the
dynamics of the two ABS branches.

(ii) Weak Landau-Zener tunneling produces avoided crossings in the
Floquet spectra, as it is the case for any generic quantum-mechanical
perturbation.

As a consequence of the above items~(i) and~(ii), the dips in the
voltage dependence of $I_{q,c}(eV/\Delta,\Phi/\Phi_0)$ and the avoided
crossings in the Floquet spectrum appear simultaneously at the same
voltage values, because they have a common origin, {\it i.e.}  quantum
superposition of the positive and negative-energy ABS manifolds, as a
result of Landau-Zener tunneling between them, see the preceding
figure~\ref{fig:supplementary-2} in
section~\ref{sec:summary-of-the-paper}.

\begin{figure*}[htb]
    \includegraphics[width=\textwidth]{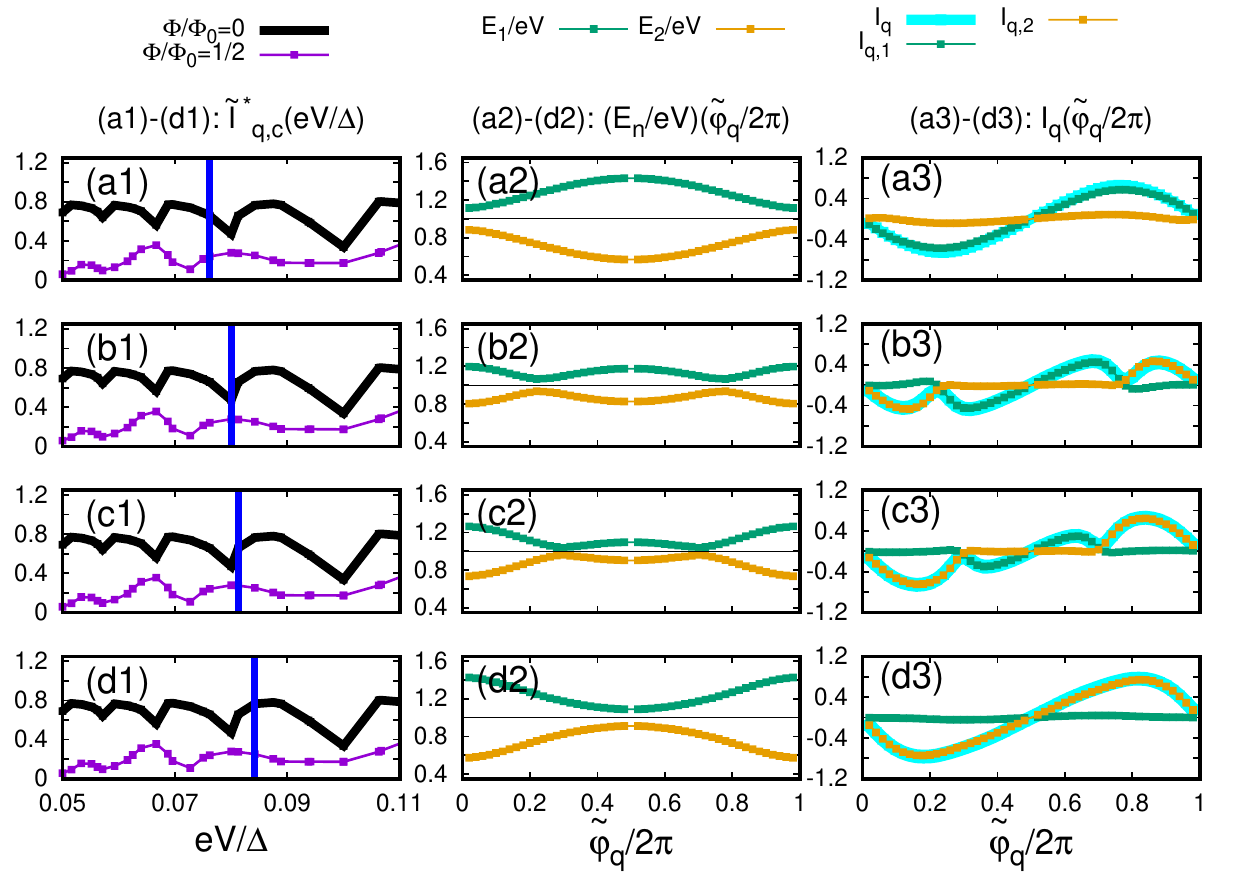}
    \caption{{\it The quartet phase sensitivity of the Floquet spectra
        and quartet current:} The figure shows the evolution of the
      reduced Floquet level energies $E_n/eV$ (see panels a2-d2) and
      the current $I_q$ (see panels a3-d3) as a function of the
      reduced quartet phase $\tilde{\varphi}_q/2\pi$. Panels a1-d1
      show the voltage values which are selected while scanning
      through a dip in $I_{q,c}^*(eV/\Delta)$ plotted as a function of
      $eV/\Delta$.
    \label{fig:dip_number_1}
  }
\end{figure*}

{\it Current carried by each Floquet state:} Now, {we discuss the
  voltage-$V$ dependence} of the currents $\tilde{I}_{1}$ and
$\tilde{I}_{2}$ carried by each Floquet state, see Eqs.~(\ref{eq:I1})
and~(\ref{eq:I2}) in the preceding section~\ref{sec:definition}.

The reduced voltage-$eV/\Delta$ dependence of
$\tilde{I}_{q,c,1}^*(eV/\Delta,\Phi/\Phi_0)$ and
$\tilde{I}_{q,c,2}^*(eV/\Delta,\Phi/\Phi_0)$ is shown in
figure~\ref{fig:low_voltage_window1}b.

At low voltage, the current is almost entirely carried by a single
Floquet state, if the voltage value is in between two avoided
crossings. The ``$+$'' and the ``$-$'' Floquet states defined by
Eqs.~(\ref{eq:E-plus}) and~(\ref{eq:E-minus}) anticross at the
$\{V_{cross,n}\}$ above, yielding alternation between ``current
carried mostly by the Floquet state $1$'', followed by ``current
carried mostly by the Floquet state $2$'', ... as the voltage is
increased, see figure~\ref{fig:low_voltage_window1}b. It is seen on
figures~\ref{fig:low_voltage_window1}a and b that the ``switching
voltages'' between $\tilde{I}_{q,c,1}^*\simeq 0$ and
$\tilde{I}_{q,c,2}^*\simeq 0$ match perfectly with the anticrossings
in the Floquet spectra, which also coincide with the deepest minima in
$\tilde{I}_{q,c}(eV/\Delta)$, see the discussion above.

{\it Generalization to the full current-phase relations: }Our previous
discussion was based on taking the maximum of the current with respect
to the gauge invariant quartet phase. Now, we focus on the Floquet
spectrum $E_n(eV/\Delta,\tilde{\varphi}_q/2\pi,\Phi/\Phi_0)$ and on
the full current-phase relations
$I_{S_c}(eV/\Delta,\tilde{\varphi}_q/2\pi,\Phi/\Phi_0)$ as a function
of the gauge-invariant quartet phase variable
$\tilde{\varphi}_q/2\pi$. Figures~\ref{fig:dip_number_1}~a1-d1 show
the critical current $\tilde{I}^*_{q,c}(eV/\Delta,\Phi/\Phi_0)$ as a
function of the reduced voltage $eV/\Delta$, the gauge-invariant
quartet phase $\tilde{\varphi}_q$ taking the value
$\tilde{\varphi}_q\equiv \tilde{\varphi}_q^*$ [see
  Eq.~(\ref{eq:quartet-critical-current})]. The
$\tilde{\varphi}_q/2\pi$-sensitivity of the Floquet spectra and the
current-phase relations are shown on panels a2-d2 and a3-d3
respectively, at the values of the reduced voltage $eV/\Delta$ which
are selected on panels a1-d1. Going from panel a1 to panel d1, we scan
voltage through one of the dips appearing at low voltage in
$I_{q,c}(eV/\Delta,\Phi/\Phi_0)$.

Figures~\ref{fig:dip_number_1}~a2-d2 reveal that {the
  dips in} $I_{q,c}(eV/\Delta,\Phi/\Phi_0)$ plotted as a function of
$eV/\Delta$ correspond to collisions between the Floquet levels
plotted as a function of $\tilde{\varphi}_q/2\pi$. Avoided crossings
appear in $E_n(eV/\Delta,\tilde{\varphi}_q/2\pi,\Phi/\Phi_0)$ plotted
as a function of $\tilde{\varphi}_q/2\pi$. Part of
figure~\ref{fig:dip_number_1} is already presented in the
Supplementary Information of the Harvard group
paper\cite{Harvard-group-experiment}. But here, panels a3-d3 show in
addition the $\tilde{\varphi}_q/2\pi$-dependence of the currents
$I_1(eV/\Delta,\tilde{\varphi}_q/2\pi,\Phi/\Phi_0)$ and
$I_2(eV/\Delta,\tilde{\varphi}_q/2\pi,\Phi/\Phi_0)$ carried by each
Floquet state, see Eqs.~(\ref{eq:I1}) and~(\ref{eq:I2}) above.

The following is deduced from figure~\ref{fig:dip_number_1}:

(i) The current
$I_{S_c}(eV/\Delta,\tilde{\varphi}_q/2\pi,\Phi/\Phi_0)$ is carried by
a single Floquet state for most of the values of
$\tilde{\varphi}_q/2\pi$, except in the immediate neighborhood of an
avoided crossing where both
$\tilde{I}_{1}(eV/\Delta,\tilde{\varphi}_q/2\pi,\Phi/\Phi_0)$ and
$\tilde{I}_{2}(eV/\Delta,\tilde{\varphi}_q/2\pi,\Phi/\Phi_0)$ have a
small contribution to
$\tilde{I}_{S_c}(eV/\Delta,\tilde{\varphi}_q/2\pi,\Phi/\Phi_0)$.

(ii) We find
$I_{S_c}(eV/\Delta,\tilde{\varphi}_q/2\pi,\Phi/\Phi_0)\simeq 0$ if the
reduced gauge-invariant quartet phase $\tilde{\varphi}_q/2\pi$ is
tuned at an avoided crossing according to the spectra on
figure~\ref{fig:dip_number_1}~a2-d2.

\begin{figure*}[htb]
  \includegraphics[width=\textwidth]{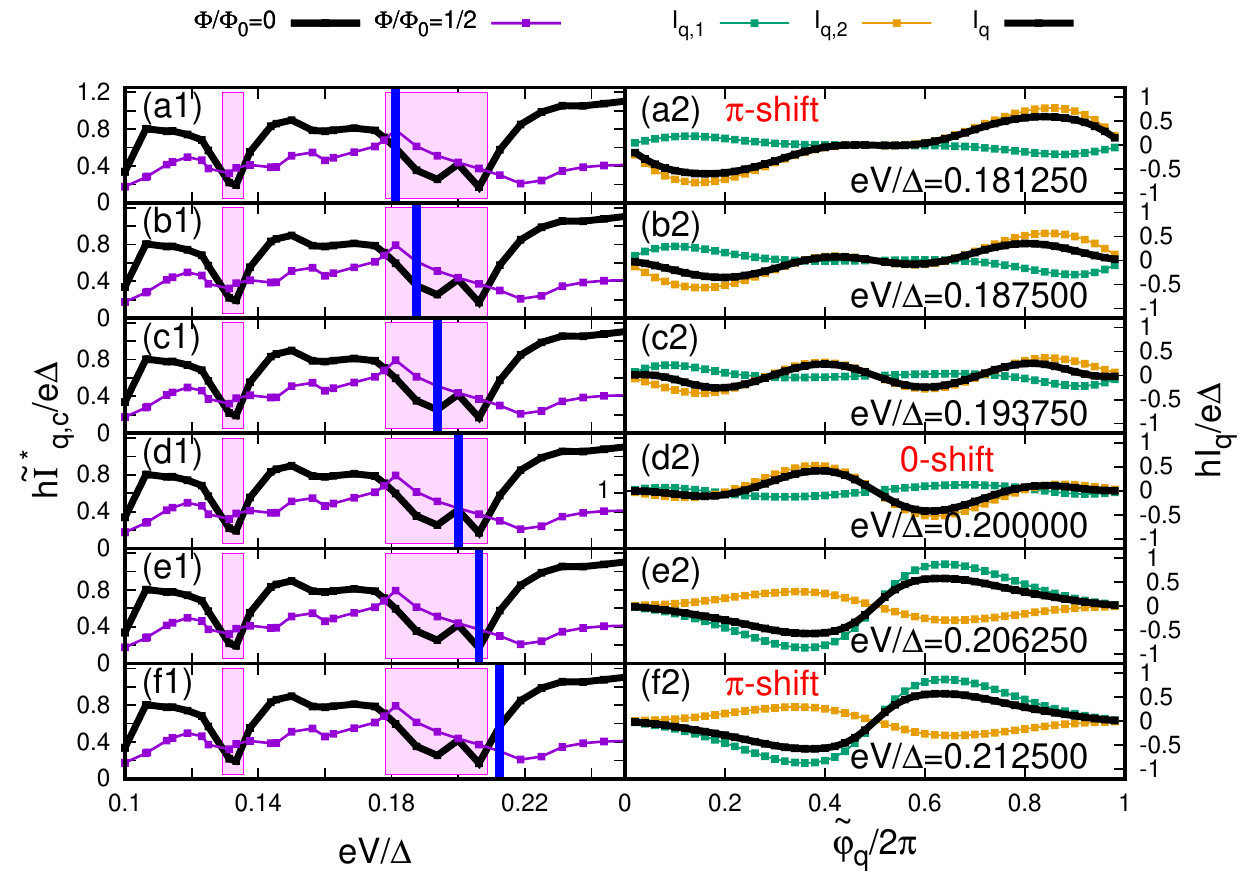}
  
  \caption{{\it The contribution of both Floquet states to the quartet
      current:} The figure shows a scan through a dip in the quartet
    critical current $I_{q,c}$, see panels~a1-f1. The quartet current
    $I_q$ and the contributions $I_{q,1}$ and $I_{q,2}$ of both
    Floquet states are shown on panels~a2-f2 as a function of the
    reduced gauge invariant quartet phase $\tilde{\varphi}_q/2\pi$.
    \label{fig:dip_number_4}
  }
\end{figure*}

It is concluded that the $\tilde{\varphi}_q$-dependence of the quartet
current confirms the link between ``repulsion in the Floquet
spectrum'' plotted as a function of the voltage~$V$ or the
gauge-invariant phase variable $\tilde{\varphi}_q$, and the ``minima
in the quartet critical current''.

{Now that we addressed hybridization between the two Floquet states,
  we consider higher values of the bias voltage on the Floquet
  populations, see subsection~\ref{sec:mechanism-generalization}.}

\subsubsection{Populating both Floquet states and the $\pi$-shift}
\label{sec:pi-shift}

{In this subsection, we discuss that a $\pi$-shifted current-phase
  relation emerges}, and how it can be interpreted as the result of
nonequilibrium Floquet populations.

Coming back to figure~\ref{fig:spectral_currents}, the evolution from
panel a2 to panel d2 across a dip in
$I_{q,c}^*(eV/\Delta,\Phi/\Phi_0)$ as a function of $eV/\Delta$
involves spectral current carried by both Floquet states if the
voltage is tuned at a minimum in $I_{q,c}^*(eV/\Delta,\Phi/\Phi_0)$,
see figure~\ref{fig:spectral_currents}~c2.

Populating both Floquet states can be realized by increasing voltage
for the considered weak Landau-Zener tunneling ({\it i.e.}
$\gamma/\Delta=0.3$ and $\Phi/\Phi_0=0$). On
figure~\ref{fig:dip_number_4}, we scan the reduced voltage $eV/\Delta$
through a dip in $\tilde{I}_{q,c}^*(eV/\Delta,\Phi/\Phi_0)$, but now
at higher $eV/\Delta$ values than on
figure~\ref{fig:dip_number_1}. The current-phase relations are shown
on figures~\ref{fig:dip_number_4}~a2-f2. A cross-over from
$\pi$-shifted current-phase relation (see
figure~\ref{fig:dip_number_4}~a2) to $0$-shift (see
figure~\ref{fig:dip_number_4}~d2) and back to $\pi$-shift (see
figure~\ref{fig:dip_number_4}~f2) is obtained as $eV/\Delta$ is
increased.

The low-bias quartet current is $\pi$-shifted, in agreement with
qualitative arguments on exchanging partners of two Cooper pairs
\cite{theorie-quartets2}, see also section V\,A in
Ref.~\onlinecite{paperI}.

The proposed interpretation of the $\pi$-$0$ and $0$-$\pi$ cross-overs
appearing at $eV/\Delta \simeq 0.2$ on figure~\ref{fig:dip_number_4}
is the following: $\pi$-shifted Josephson relation was obtained in a
superconductor-normal metal-superconductor ($SNS$) Josephson weak
link, originating from injection of nonequilibrium quasiparticle
populations from two attached normal leads \cite{vanWees}. This
$\pi$-shifted current-phase relation can be interpreted by noting that
the two ABS at opposite energies carry opposite currents. A change of
sign in the current-phase relation is obtained if the positive-energy
ABS is mostly populated. This is why we relate the $\pi$-$0$ and the
$0$-$\pi$ shifts of $I_{q,c}$ to the nonequilibrium Floquet
populations produced for these relatively large values of the reduced
voltage $eV/\Delta$.

The $\sim - \sin(2\varphi_q)$ current-phase relation
appearing at the $\pi$-$0$ cross-over on panel c2 meets physical
expectations regarding emergence of a second-order harmonics of the
current-phase relation once the first-order harmonics changes sign.


\begin{figure*}[htb]
  \includegraphics[width=\textwidth]{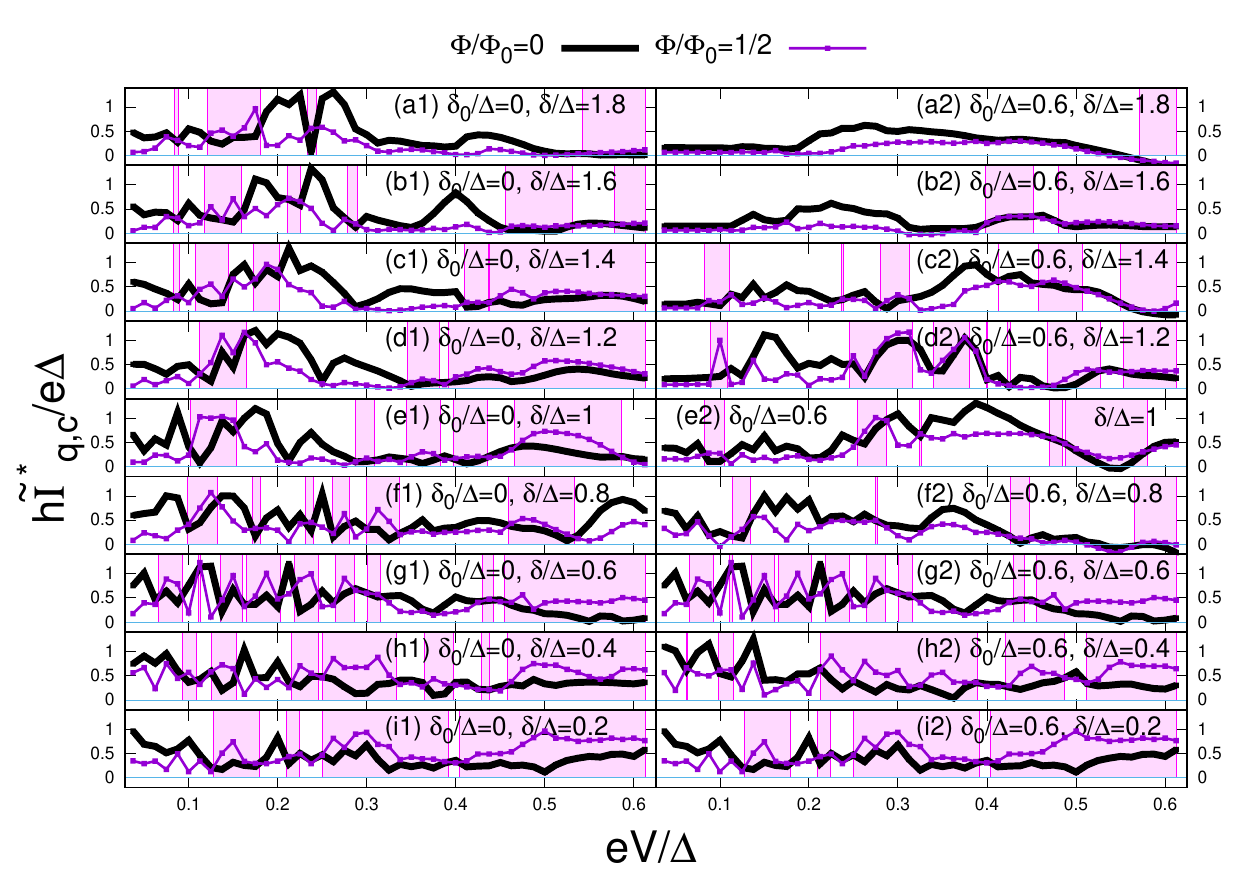}
  \caption{{\it Inversion for a multilevel quantum dot:} The figure
    shows $\tilde{I}_{q,c}^*$ as a function of reduced voltage
    $eV/\Delta$ for the multilevel quantum dot model with
    $\Phi/\Phi_0=0$ (black lines) and $\Phi/\Phi_0=1/2$ (magenta
    lines). The magenta shaded region corresponds to the inversion.
    The parameter $\delta_0/\Delta=0$ is used on panels a1-i1, with
    $\delta/\Delta$ ranging from $1.8$ (panel a1) to $0.2$ (panel i1).
    The parameter $\delta_0/\Delta=0.6$ is used on panels a2-i2, and
    $\delta/\Delta$ is from $1.8$ (panel a2) to $0.2$ (panel i2).
        \label{fig:inversion-multilevel-quantum-dot-1}
    }
\end{figure*}

\subsection{Conclusion on this section}
\label{sec:summary}

To summarize, the inversion in $I_{q,c}(V,\Phi/\Phi_0)$ between
$\Phi/\Phi_0=0$ and $\Phi/\Phi_0=1/2$, {\it i.e.} $I_{q,c}(V,0) <
I_{q,c}(V,1/2)$ emerges in our quantum dot model calculations. The
mechanism was anticipated in the above
section~\ref{sec:summary-of-the-paper} and our numerical calculations
for the voltage-$V$ and the quartet phase
$\tilde{\varphi}_q$-sensitivity of the quartet current confirmed the
proposed mechanism. Namely, in the limit of weak Landau-Zener
tunneling and with a single level quantum dot, the inversion was
interpreted as reduction in the quartet current in the vicinity of the
crossings in the Floquet spectrum.

In addition, we obtained evidence for $\pi$-shift in the quartet
current-phase relation in a narrow window of the reduced voltage
$eV/\Delta$. This numerical result was interpreted as being a
consequence of nontrivial Floquet populations.

\section{Robustness of the inversion}
\label{sec:num2}

Now, we investigate robustness of the inversion against strong
Landau-Zener tunneling and many levels in the quantum dot.

In section~III\,A of the Supplemental Material, we show that the
connection between the extrema in the Floquet spectrum and the minima
in the quartet critical current (both being plotted as a function of
reduced voltage $eV/\Delta$) holds also for strong Landau-Zener
tunneling with $\gamma/\Delta=-0.25$ (see the previous
section~\ref{sec:R}). Next, section~III\,B of the Supplemental
Material presents a scan from $\gamma/\Delta=-0.25$ to
$\gamma/\Delta=0.3$, and provides evidence for inversion in this range
of $\gamma/\Delta$.

Now, we show that inversion
$I_{q,c}(eV/\Delta,0)<I_{q,c}(eV/\Delta,1/2)$ appears generically for
the multilevel quantum dot presented in the previous
section~\ref{sec:num1}, specialized to the equally
spaced energy levels:
\begin{equation}
\epsilon_n=n \delta + \delta_0
,
\end{equation}
with $n$ an integer. An estimate for the number of energy levels
within the gap window is $2\Delta/\delta$.

Figures~\ref{fig:inversion-multilevel-quantum-dot-1}~a1-i1 and
figure~\ref{fig:inversion-multilevel-quantum-dot-1}~a2-i2 correspond
to $\delta_0/\Delta=0$ and $\delta_0/\Delta=0.6$ respectively, with
$\delta/\Delta$ ranging from $\delta/\Delta=1.8$ (panels a1 and a2) to
$\delta/\Delta=0.2$ (panels i1 and i2). Panels i1 and i2 coincide with
to each other, because $(\delta_0/\Delta,\delta/\Delta)=(0,0.2)$ and
$(\delta_0/\Delta,\delta/\Delta)=(0.6,0.2)$ produce the same spectrum
of the quantum dot energy levels.

It is concluded from
figures~\ref{fig:inversion-multilevel-quantum-dot-1}~a1-i1 and
figures~\ref{fig:inversion-multilevel-quantum-dot-1}~a2-i2 that
crossing-over from $\delta/\Delta=1.8$ larger than unity on
figures~\ref{fig:inversion-multilevel-quantum-dot-1}~a1-a2 (typically
with zero of a single energy level in the gap window) to
$\delta/\Delta=0.2$ on
figures~\ref{fig:inversion-multilevel-quantum-dot-1}~i1-i2 (with~$\sim
10$ energy levels in the gap window) generically implies emergence of
inversion. It is concluded from
figure~\ref{fig:inversion-multilevel-quantum-dot-1} that the inversion
is favored upon increasing the number of levels on the quantum dot, in
comparison with a single level quantum dot.

\section{Conclusions}
\label{sec:conclusions}

This paper addressed a four-terminal $(S_a,S_b,S_{c,1},S_{c,2})$
quantum dot-Josephson junction biased at $(V,-V,0,0)$ on the quartet
line (see the device on figure~\ref{fig:device}). The quartet critical
current $I_{q,c}(eV/\Delta,\Phi/\Phi_0)$ is parameterized by both the
reduced voltage $eV/\Delta$ and the reduced flux $\Phi/\Phi_0$
piercing through the loop. It turns out that the recent Harvard group
experiment \cite{Harvard-group-experiment} observes ``inversion''
between $\Phi/\Phi_0=0$ and $\Phi/\Phi_0=1/2$, namely
$I_{q,c}(eV/\Delta,\Phi/\Phi_0)$ can be larger at $\Phi/\Phi_0=1/2$
than at $\Phi/\Phi_0=0$. This experimental result is against the naive
expectation that destructive interference should reduce the quartet
critical current at $\Phi/\Phi_0=1/2$ with respect to $\Phi/\Phi_0=0$.

We addressed in this paper~II how inversion can be produced at finite
bias voltage~$V$ in a simple 0D quantum dot device. The ``Floquet
mechanism'' for the inversion tuned by the voltage~$V$ is simple in
the limit of weak Landau-Zener tunneling. First, in the absence of
Landau-Zener tunneling between the two ABS manifolds, the classical
Floquet spectrum shows nonavoided crossings as a function of the
reduced voltage $eV/\Delta$. Second, the rate of Landau-Zener
tunneling increases from zero as $eV/\Delta$ is increased. This yields
opening of gaps in the Floquet spectrum, which makes the crossings
between the Floquet levels become avoided. The quantum mechanical
effects of weak Landau-Zener tunneling are important only if the bias
voltage is close to avoided crossings in the Floquet spectra.
Landau-Zener tunneling produces hybridization between the two Floquet
states and a reduction of the quartet critical current
$I_{q,c}(eV/\Delta,\Phi/\Phi_0)$, due to the time-dependent dynamical
quantum superpositions of the two ABS which carry opposite
currents. In certain voltage windows, the reduction in
$I_{q,c}(eV/\Delta,\Phi/\Phi_0)$ at $\Phi/\Phi_0=0$ is such as to
produce inversion with $\Phi/\Phi_0=1/2$. In addition, we demonstrated
that nontrivial populations of the two Floquet states are produced at
larger voltage, which yields change of sign in the relation between
the quartet current and the gauge-invariant phase variable.

Finally, our results suggest that the inversion is generic since it
holds also for strong Landau-Zener tunneling and for a multilevel
quantum dot, which is encouraging with respect to providing mechanisms
for the recent Harvard group experiment
\cite{Harvard-group-experiment}. In the forthcoming paper~III of the
series, we will evaluate the voltage-$V$ sensitivity for the more
realistic ``2D metal beam splitter'' proposed in the previous paper~I
\cite{paperI} (instead of the 0D quantum dot of this paper~II).

\section*{Acknowledgements}

The authors acknowledge the very stimulating collaboration with the
Harvard group (K. Huang, Y. Ronen and P. Kim) on the interpretation of
their experiment, on the identification of the most relevant numerical
results and on the way to present them. The authors acknowledge useful
discussions with D. Feinberg. R.M. wishes to thank R. Danneau for
fruitful discussions on the way to present the results. R.M. thanks
the Infrastructure de Calcul Intensif et de Donn\'ees (GRICAD) for use
of the resources of the M\'esocentre de Calcul Intensif de
l’Universit\'e Grenoble-Alpes (CIMENT).


\end{document}